\documentstyle[12pt]{article}

\def\hybrid{\topmargin -20pt	\oddsidemargin 0pt
	\headheight 0pt	\headsep 0pt
        \textwidth 6.35in
        \textheight 9.65in
	\marginparwidth .875in
	\parskip 5pt plus 1pt	\jot = 1.5ex}

\def\baselinestretch{1.2}

\catcode`\@=11

\def\marginnote#1{}
\newcount\hour
\newcount\minute
\newtoks\amorpm
\hour=\time\divide\hour by60
\minute=\time{\multiply\hour by60 \global\advance\minute by-\hour}
\edef\standardtime{{\ifnum\hour<12 \global\amorpm={am}%
	\else\global\amorpm={pm}\advance\hour by-12 \fi
	\ifnum\hour=0 \hour=12 \fi
	\number\hour:\ifnum\minute<10 0\fi\number\minute\the\amorpm}}
\edef\militarytime{\number\hour:\ifnum\minute<10 0\fi\number\minute}
\def\draftlabel#1{{\@bsphack\if@filesw {\let\thepage\relax
   \xdef\@gtempa{\write\@auxout{\string
      \newlabel{#1}{{\@currentlabel}{\thepage}}}}}\@gtempa
   \if@nobreak \ifvmode\nobreak\fi\fi\fi\@esphack}
	\gdef\@eqnlabel{#1}}
\def\@eqnlabel{}
\def\@vacuum{}
\def\draftmarginnote#1{\marginpar{\raggedright\scriptsize\tt#1}}

\def\draft{\oddsidemargin -.2truein
	\def\@oddfoot{\sl preliminary draft \hfil
	\rm\thepage\hfil\sl\today\quad\militarytime}
	\let\@evenfoot\@oddfoot	\overfullrule 3pt
	\let\label=\draftlabel
	\let\marginnote=\draftmarginnote
   \def\@eqnnum{(\theequation)\rlap{\kern\marginparsep\tt\@eqnlabel}%
\global\let\@eqnlabel\@vacuum}  }

\def\preprint{\twocolumn\sloppy\flushbottom\parindent 2em
	\leftmargini 2em\leftmarginv .5em\leftmarginvi .5em
	\oddsidemargin -.5in	\evensidemargin -.5in
	\columnsep .4in	\footheight 0pt
	\textwidth 10.in	\topmargin  -.4in
	\headheight 12pt \topskip .4in
	\textheight 6.9in \footskip 0pt
	\def\@oddhead{\thepage\hfil\addtocounter{page}{1}\thepage}
	\let\@evenhead\@oddhead	\def\@oddfoot{}	\def\@evenfoot{} }

\def\numberbysection{\@addtoreset{equation}{section}
	\def\theequation{\thesection.\arabic{equation}}}

\def\underline#1{\relax\ifmmode\@@underline#1\else
	$\@@underline{\hbox{#1}}$\relax\fi}

\def\titlepage{\@restonecolfalse\if@twocolumn\@restonecoltrue
\onecolumn
     \else \newpage \fi \thispagestyle{empty}\c@page\z@
	\def\thefootnote{\fnsymbol{footnote}} }

\def\endtitlepage{\if@restonecol\twocolumn \else \newpage \fi
	\def\thefootnote{\arabic{footnote}}
	\setcounter{footnote}{0}}  

\catcode`@=12
\relax

\def\np{Nucl. Phys. \/}
\def\pl{Phys. Lett. \/}
\def\figcap{\section*{Figure Captions\markboth
	{FIGURECAPTIONS}{FIGURECAPTIONS}}\list
	{Figure \arabic{enumi}:\hfill}{\settowidth\labelwidth{Figure 999:}
	\leftmargin\labelwidth
	\advance\leftmargin\labelsep\usecounter{enumi}}}
 \relax
\def\tablecap{\section*{Table Captions\markboth
	{TABLECAPTIONS}{TABLECAPTIONS}}\list
	{Table \arabic{enumi}:\hfill}{\settowidth\labelwidth{Table 999:}
	\leftmargin\labelwidth
	\advance\leftmargin\labelsep\usecounter{enumi}}}
 \relax
\def\reflist{\section*{References\markboth
	{REFLIST}{REFLIST}}\list
	{[\arabic{enumi}]\hfill}{\settowidth\labelwidth{[999]}
	\leftmargin\labelwidth
	\advance\leftmargin\labelsep\usecounter{enumi}}}
 \relax
\makeatletter
\newcounter{pubctr}
\def\publist{\@ifnextchar[{\@publist}{\@@publist}}
\def\@publist[#1]{\list
	{[\arabic{pubctr}]\hfill}{\settowidth\labelwidth{[999]}
	\leftmargin\labelwidth
	\advance\leftmargin\labelsep
	\@nmbrlisttrue\def\@listctr{pubctr}
	\setcounter{pubctr}{#1}\addtocounter{pubctr}{-1}}}
\def\@@publist{\list
	{[\arabic{pubctr}]\hfill}{\settowidth\labelwidth{[999]}
	\leftmargin\labelwidth
	\advance\leftmargin\labelsep
	\@nmbrlisttrue\def\@listctr{pubctr}}}
 \relax
\makeatother
\newskip\humongous \humongous=0pt plus 1000pt minus 1000pt

\newif\ifdtup

\relax
\hybrid

\def\thefootnote{\fnsymbol{footnote}}
\def\be{\begin{equation}}
\def\ee{\end{equation}}
\def\ba{\begin{eqnarray}}
\def\ea{\end{eqnarray}}
\def\a{\alpha}
\def\b{\beta}
\def\g{\gamma}
\def\p{\partial}
\def\pb{\bar \partial}
\def\th{\vartheta}

\def\l{\lambda}
\def\square{\hbox{{$\sqcup$}\llap{$\sqcap$}}}

\def\im{{\rm Im}\tau}
\def\R{{\cal{R}}}
\def\Q{{\cal{Q}}}
\def\RR{{\rm I\!R}}
\def\n{\nabla}
\def\pq{\bar {\cal P}}
\def\rt{{\cal R}}
\font\fivesans=cmss10 at 4.61pt
\font\sevensans=cmss10 at 6.81pt
\font\tensans=cmss10
\newfam\sansfam
\textfont\sansfam=\tensans\scriptfont\sansfam=
\sevensans\scriptscriptfont
\sansfam=\fivesans
\def\sans{\fam\sansfam\tensans}

\def\Z{{\mathchoice
{\hbox{$\sans\textstyle Z\kern-0.4em Z$}}
{\hbox{$\sans\textstyle Z\kern-0.4em Z$}}
{\hbox{$\sans\scriptstyle Z\kern-0.3em Z$}}
{\hbox{$\sans\scriptscriptstyle Z\kern-0.2em Z$}}}}
\newlength{\boxsize} 

\begin{document}
\renewcommand{\theequation}{\thesection.\arabic{equation}}
\newcommand{\beq}{\begin{equation}}
\newcommand{\eeq}[1]{\label{#1}\end{equation}}
\newcommand{\ber}{\begin{eqnarray}}
\newcommand{\eer}[1]{\label{#1}\end{eqnarray}}
\begin{titlepage}
\begin{center}

\hfill CERN-TH/95-171\\
\hfill LPTENS-95/28\\
\hfill hep-th/9508078\\

\vskip .2in

{\large \bf Infrared behavior of Closed Superstrings in Strong
Magnetic and Gravitational Fields}
\vskip .4in

{\bf Elias Kiritsis and Costas Kounnas\footnote{On leave from Ecole
Normale Sup\'erieure, 24 rue Lhomond, F-75231, Paris, Cedex 05,
FRANCE.}}\\
\vskip
 .3in

{\em Theory Division, CERN,\\ CH-1211,
Geneva 23, SWITZERLAND} \footnote{e-mail addresses:
KIRITSIS,KOUNNAS@NXTH04.CERN.CH}\\

\vskip .3in

\end{center}

\vskip .2in

\begin{center} {\bf ABSTRACT }
 \end{center}
\begin{quotation}\noindent
A large class of four-dimensional supersymmetric ground states of
closed superstrings with a non-zero mass gap are constructed.
For such ground states we turn on chromo-magnetic fields as well as
curvature.
The exact spectrum as function of the chromo-magnetic fields and
curvature is derived.
We examine the behavior of the spectrum, and find that there is a
maximal value for the magnetic field $H_{\rm max}\sim M_{\rm
planck}^2$. At
this value all states that couple to the magnetic field become
infinitely massive and decouple.
We also find tachyonic
instabilities for
strong background fields of the order ${\cal O}(\mu M_{\rm planck})$
where $\mu$ is the mass gap of the theory. Unlike the field theory
case, we find that such ground states become stable again for
magnetic fields
of the order ${\cal O}(M^2_{\rm planck})$.
The implications of these results are discussed.

\end{quotation}
\vskip 2.cm
CERN-TH/95-171 \\
August 1995\\
\end{titlepage}
\vfill
\eject
\def\baselinestretch{1.2}
\baselineskip 16 pt
\noindent
\section{Introduction}
\setcounter{equation}{0}

In  four-dimensional Heterotic or type II  Superstrings it is
possible in principle to understand the response of the theory  to
non-zero  gauge or  gravitational field  backgrounds including
quantum corrections.
This problem is difficult in its full generality since
we are working in a first quantized framework. In certain special
cases, however, there is an underlying 2-d superconformal theory
which
is well understood and which describes exactly (via marginal
deformations) the turning-on of non-trivial
gauge and gravitational backgrounds.
This exact description goes beyond the linearized approximation.
In such cases, the spectrum can be calculated exactly, and it can
provide
interesting clues about the physics of the theory.

In field theory (excluding gravity) the
energy shifts of a state due to the magnetic field have been
investigated long ago \cite{field,qcd,stand}. The classical field
theory
formula for the energy of a state of a spin $S$, mass $M$ and charge
$e$ in a magnetic field
$H$ pointing in the third direction is:
\be
E^2=p_3^2+M^2+|eH|(2n+1-gS)\label{class}
\ee
were $g=1/S$ for minimally coupled states and $n=1,2,...$ labels the
Landau levels.
It is obvious from (\ref{class}) that minimally coupled particles
cannot become tachyonic, so the theory is stable.
For non-minimally coupled particles, however, the factor $2n+1-gS$
can become negative and instabilities thus appear.
For example, in non-abelian gauge theories, there are particles which
are not minimally coupled. In the standard model, the $W^{\pm}$
bosons
have $g=2$ and $S=\pm 1$. From (\ref{class}) we obtain that the
spontaneously broken phase in the standard model is thus unstable for
 magnetic fields that satisfy \cite{qcd,stand}
\be
|H|\geq {M_{W}^2\over |e|}  \label{sta}
\ee
A phase transition has to occur by a condensation of $W$ bosons, most
probably to a phase where the magnetic field is confined (localized)
in a tube, \cite{stand}.
This behavior should be contrasted to the constant electric field
case where
there is particle production \cite{sch} for any non-zero value of the
electric field, but the vacuum is stable (although the particle
emission tends to decrease the electric field).

The instabilities present for constant magnetic fields are still
present in general for slowly varying (long range) magnetic fields.
For a non-abelian gauge theory in the unbroken phase, since the mass
gap
is classically zero, we deduce from (\ref{class}) that the trivial
vacuum
($A_{\mu}^{a}=0$) is unstable even for infinitesimally small
chromo-magnetic
fields. This provides already an indication at the classical level
that the trivial vacuum is not the correct vacuum in an unbroken
non-abelian gauge theory. We know however, that such a theory
acquires a non-perturbative mass gap, $\Lambda^2\sim \mu^2
\exp[-16\pi^2 b_{0}/g^2]$ where $g$ is the non-abelian gauge
coupling.
If in such a
theory one managed to create a chromo-magnetic field then there would
again appear an instability and the theory would again confine the
field in a flux tube.

In string theory, non-minimal gauge couplings are present not only in
the massless sector but also in the massive (stringy) sectors
\cite{fpt}.
Thus one would expect similar instabilities.
Since in  string theory there are states with arbitrary large values
of spin
and one can naively expect that if  $g$ does not decrease fast enough
with the spin
(as is the case in open strings where $g=2$ \cite{fpt}) then for
states with
large spin an arbitrarily small magnetic field would destabilize the
theory. This behavior would imply that the trivial vacuum is
unstable.
This does not happen however since the masses of particles with spin
also become large when the spin gets large.

The spectrum of open bosonic strings in constant magnetic fields was
derived in \cite{acny}. Open bosonic strings however, contain
tachyons even in the absence
of background fields. It is thus more interesting to investigate
open superstrings which are tachyon-free.
This was done in  \cite{fp}.
It was found that for weak magnetic fields the field theory formula
(\ref{class}) is obtained, and there are similar instabilities.

In closed superstring theory however, one is forced to include the
effects of gravity.
A constant magnetic field for example carries energy, thus, the space
cannot remain flat anymore.
The interesting question in this context is, to what extend, the
gravitational backreaction changes the behavior seen in field theory
and open string theory.

Such questions can have potential interesting applications in string
cosmology
since long range magnetic fields can be produced at early stages in
the history of the universe where field theoretic behavior can be
quite different from the stringy one.

The first example of an exact electromagnetic solution to closed
string theory
was described in \cite{bas}.
The solution included both an electric and magnetic field
(corresponding to the electrovac solution of supergravity).
In \cite{bk} another exact closed string solution was presented
(among others)
which corresponded to a Dirac monopole over $S^2$.
More recently, several other magnetic solutions were presented
corresponding to
localized \cite{melvin} or covariantly constant magnetic fields
\cite{rt}.
The spectrum of these magnetic solutions seems to have a different
behavior
as a function of the magnetic field, compared to the situation
treated in this paper. The reason for this is that \cite{rt}
considered magnetic solutions where the gravitational backreaction
produces a non-static metric.
``Internal" magnetic fields of the type described in \cite{bk} were
also considered recently \cite{b} in order to break spacetime
supersymmetry.

Here we will study  the effects of  covariantly  constant
(chromo)magnetic fields, $H^a_i=\epsilon^{ijk}F_{jk}^a$ and constant
curvature
${\cal R}^{il}=\epsilon^{ijk}\epsilon^{lmn}{\cal R}_{jm,kn}$, in
four-dimensional closed superstrings.
The relevant framework was developed in \cite{kk} where ground
states were
found, with a continuous (almost constant) magnetic field in a weakly
curved space.
We will describe in this paper the detailed construction of such
ground states
and we will eventually study their behavior.

In the heterotic string (where the left moving
sector is N=1 supersymmetric) the part of the $\sigma$-model action
which corresponds to a gauge field background $A^{a}_{\mu}(x)$
is
\be
V=(A^{a}_{\mu}(x)\partial x^{\mu}+F^{a}_{ij}(x)\psi^{i}\psi^{j})\bar
J^{a}
\label{vertex}
\ee
where $F^{a}_{ij}$ is the field strength of $A^{a}_{\mu}$ with
tangent
space indices, eg. $F^{a}_{ij}=e^{\mu}_{i}e^{\nu}_{j}F^{a}_{\mu\nu}$
with $e^{\mu}_{i}$ being the inverse vielbein, and $\psi^{i}$ are
left-moving world-sheet fermions with a normalized kinetic term.
$\bar J^{a}$ is a right moving affine current.

Consider a string ground state with a flat non-compact (euclidean)
spacetime ($\RR^{4}$).
The simplest case to consider is that of a constant magnetic field,
$H^a_i=\epsilon^{ijk}F_{jk}^a$.
Then the relevant vertex operator (\ref{vertex}) becomes
\be
V_{flat}=F_{ij}^a({1\over 2}x^{i} \partial
x^{j}+\psi^{i}\psi^{j})\bar
J^{a}
\label{vertexflat}
\ee
This vertex operator however, cannot be used to turn on the magnetic
field
since it is not marginal (when $F_{ij}^{a}$ is constant). In other
words, a constant magnetic in flat
space
does not satisfy the string equations of motion, in particular the
ones
associated with the gravitational sector.

A way to bypass this problem we need to switch on more background
fields.
In \cite{kk} we achieved this in two steps.
First, we found an exact string ground state in which $\RR^{4}$ is
replaced by
$\RR\times S^{3}$. The $\RR$ part corresponds to free boson with
background charge
$Q=1/\sqrt{k+2}$ while the $S^{3}$ part corresponds to an $SU(2)_{k}$
WZW model. For any (positive integer) k, the combined central charge
is equal to that of $\RR^{4}$. For large $k$, this background has a
linear dilaton in the $x^{0}$ direction as well as an
$SO(3)$-symmetric antisymmetric tensor on $S^{3}$, while the metric
is
the standard round metric on $S^{3}$ with constant
curvature.
On this space, there is an exactly marginal vertex operator for a
magnetic field which is
\be
V_{m}=H(J^{3}+\psi^{1}\psi^{2})\bar J^{a}\label{magnet}
\ee
Here, $J^{3}$ is the left-moving current of the $SU(2)_{k}$ WZW
model.
$V_{m}$ contains the only linear combination of $J^{3}$ and
$\psi^{1}\psi^{2}$
that does not break the N=1 local supersymmetry.
The exact marginality of this vertex operator is obvious since it is
a product of a left times a right abelian current.
This operator is unique up to an $SU(2)_{L}$ rotation.

We can observe  that this vertex operator provides a well defined
analog
of $V_{flat}$ in eq. (\ref{vertex}) by looking at the large $k$
limit.
We will write the SU(2) group element as
$g=\exp[i{\vec\sigma}\cdot{\vec x}/2]$ in which case
$J^{i}=kTr[\sigma^{i}
g^{-1}\p g]=ik(\p x^{i}+\epsilon^{ijk}x_j\p x_k+{\cal{O}}(|x|^3))$.
In the flat limit the first term corresponds to a constant gauge
field
and thus pure gauge so the only relevant term is the second one which
corresponds to constant magnetic field in flat space.
The fact  $\pi_{2}(S^{3})=0$ explains in a different way why there is
no quantization condition on $H$.
THis magnetic field background break spacetime supersymmetry as
usually expected.
It evades one of the assumptions of the Banks-Dixon theorem \cite{BD}
since it involves a vertex operator from non-compact 4-d spacetime.

There is another exactly marginal perturbation in the background
above
that turns on fields in the gravitational sector.
The relevant perturbation is
\be
V_{grav}=\R (J^{3}+\psi^{1}\psi^{2})\bar J^{3}
\label{gravi}\ee
This perturbation modifies the metric, antisymmetric tensor and
dilaton \cite{kk}.
For type II strings the relevant perturbation is
\be
V_{grav}^{II}=\R(J^3+\psi^{1}\psi^{2})(\bar J^{3}+\bar\psi^{1}\bar
\psi^{2})
\ee

The space we are using, $\RR\times S^3$ is such that the spectrum has
a mass gap
$\mu^2$.
In particular all gauge symmetries are broken spontaneously. This
breaking
however is not the standard breaking due to a constant expectation
value of a scalar but due to non-trivial expectation values of the
fields in the universal sector (graviton, antisymmetric tensor and
dilaton).

In subsequent sections we derive the exact spectrum of closed string
ground states in the presence of magnetic and gravitational fields
generated
by (\ref{magnet}), (\ref{gravi}).
An interesting first result is that such magnetic fields in closed
strings
cannot become larger than a maximal value
\be
H_{\rm max}={M^2_{\rm planck}\over \sqrt{2}}
\ee
where $M_{\rm planck}=M_{\rm string}/g_{\rm string}$, $M_{\rm
string}=1/\sqrt{\alpha'}$ and $g_{\rm
string}$ is the string coupling constant.
This is reminiscent of the appearance of a (finite) maximal electric
field in open superstrings \cite{bp}.
There, the maximal electric field is associated with the value at
which the pair-production rate per unit volume becomes infinite.
Here at $H=H_{\rm max}$ all states that couple to the magnetic field
(i.e. having non-zero charge and/or angular momentum) become
infinitely massive.
This phenomenon is similar to the limit ${\rm Im} U\to \infty$ in a
2-d torroidal
CFT. It is thus a boundary of the magnetic field moduli space.

Since the theories we consider have a mass gap, they are stable for
small magnetic fields. As we keep increasing the magnetic field we
find an instability for $|H|\geq H^{\rm crit}_{\rm lower}$ whose
origin is similar to the field theoretic one, namely
states with helicity one and non-minimal coupling become tachyonic.
Unlike field theory though (where $H^{\rm crit}_{\rm lower}\sim
\mu^2$)
here we find
\be
H^{\rm crit}_{\rm lower}\sim \mu M_{\rm planck}
\ee
The reason for the different behavior can be traced to the way the
gauge symmetry is broken in field theory versus our string ground
states.
In field theory higher helicity particles with non-minimal couplings
(like charged gauge bosons) get a mass from the expectation value of
a charged scalar (Higgs field). In the string ground states we
consider, the gauge symmetries are broken by non-trivial expectations
values in the generic sector
(graviton, antisymmetric tensor, dilaton).

Another major difference with field theory behavior is the
following:
In field theory, for $H\geq H^{\rm crit}_{\rm lower}$ the theory is
unstable
for arbitrarily high magnetic fields.
In closed string theory we find that theory generically becomes
stable again for
\be
H^{\rm crit}_{\rm upper}\leq |H|\leq H_{\rm max}\label{up}
\ee
with
\be
H_{\rm max}-H^{\rm crit}_{\rm upper}\sim \mu M_{\rm planck}
\ee
One might think that this region is irrelevant since the theory
undergoes a phase transition already for smaller magnetic fields.
One however, could imagine a situation where a strong localized
magnetic
field (which does not induce an instability) starts spreading out in
space
to become a long range field in the region described by (\ref{up}).
The geometry of spacetime, although deformed by the presence of the
magnetic field, remains smooth (free of singularities) for the whole
range $0\leq |H|
\leq H_{\rm max}$.

Similar remarks apply to the gravitational perturbation
(\ref{gravi}).
There, we can describe this perturbation with a modulus $0\leq
\lambda\leq 1$ so that
$\lambda=1$ corresponds to the round three-sphere geometry.
For $\lambda\not= 1$ the sphere is squashed as can be seen from the
expression
of the scalar curvature (in Euler angles):
\be
R_{\rm scalar} ={8\over
k}{-1+5\lambda^2-\lambda^4+2H^2\lambda^2+(1-\lambda^4)\cos\beta\over
(
1+\lambda^2+(\lambda^2-1)\cos\beta)^2}
\ee
where we have also included the effects of a magnetic field $H$.
We note that again the geometry is smooth for $\lambda\not= 0$, while
it becomes singular at $\lambda=0$. This singularity corresponds to
the classical
singularity of the $SU(2)_{k}/U(1)$ coset model \cite{gk} which is
known to be absent from the corresponding CFT.
Simply, at $\lambda=0$ one dimension decompactifies.
Here again we find an instability for
\be
0\leq \lambda_{\rm lower}\leq \lambda \leq \lambda_{\rm upper}\leq 1
\ee

The structure of this paper is as follows.
In section 2 we describe how to construct, for any given 4-d flat
space ground state of the superstring, another ground state in curved
4-d space, with similar matter spectrum, but with a non-zero mass gap
$\mu^2$.
In section 3 we give the $\sigma$-model description of turning on
non-trivial
magnetic fields
and curvature in the ground states described in section 2.
In section 4 we give the CFT description of such magnetic
fields.
In section 5 we discuss the flat space limit when the mass gap
$\mu^2\to 0$.
Finally section 6 contain our conclusions and further directions.

\section{Construction of curved  ${W_{k}}\otimes K$ string solutions}
\setcounter{equation}{0}

Our aim is to replace the Euclidean four-dimensional flat space
solution $\RR^4\otimes K$  by a curved space solution $W_{k}\otimes
K$, where we replace the four non-compact (super)coordinates of flat
space by the (super)coordinates of the $SU(2)_k\otimes \RR_Q$ theory.
Three of the coordinates describe $SU(2)_k$ (the three-dimensional
sphere) and the fourth is a  flat coordinate with non zero background
charge $Q=1/\sqrt{k+2}$. The relation among the level $k$ (a
non-negative integer) and the background charge $Q$ is such that the
left and right central charges remains the same as in  $\RR^4$  for
any value of $k$. This give us the possibility to keep unchanged the
internal superconformal theory $K$.

We will show below that the replacement $\RR^4\rightarrow W_k$ can be
done in a universal way and without reducing the number of spacetime
supersymmetries in almost all interesting cases with non-maximal
number of supersymmetries. In the case of  maximal supersymmetry,
($N=4$ in heterotic and $N=8$ in type II), this  replacement,
although is still universal, reduces by a factor of two the number of
space time supersymmetries.  This reduction is unavoidable and  due
to the fact that only half of the constant killing spinors of $\RR^4$
remain covariantly constant  in  the $W_k$ space
\cite{wormclas,worm,bil}. The non-zero
torsion and  dilaton are responsible for this.

Let us start first with the case of maximal supersymmetry in flat
space. Following \cite{BD}, the world sheet super-current(s) of the
internal $K$ theory can be always constructed in terms of six free
bosons compactified on a torus and six free fermions\footnote{There
are some exotic cases though \cite{cp}, which will not be discussed
here.}; the internal  fermions, the $\RR^4$ fermions and the $\beta,
\gamma$ ghosts of supereparametrizations  must have identical
boundary conditions (periodic or antiperiodic) in all non-trivial
worldsheet cycles. In type II, this global restriction  must be
respected separately for the left- and right-moving fermions and the
$\beta, \gamma$ ghosts. Also, the left and right  momenta for the
compactified bosons in $K$ must form necessarily a self-dual
lorentzian lattice.

Both $\RR^4$ and $W_k$ are $N=4$ ${\hat c}=4$ superconformal
theories.
In both cases the $SU(2)_1$  $N=4$ currents $S^i$, $i=1,2,3$ are
constructed in terms of word sheet fermions $\psi_{0}$, $\psi_i$,
$i=1,2,3$
\be
S^{i}= \frac{1}{2}\left(\psi_{0}\psi^{i}+
\frac{1}{2}\epsilon^{ijl}\psi_{j}\psi_{l}\right)\ .
\ee

Observe that only the three self-dual currents appear in the
algebra.
It order to specify better the difference among the $\RR^4$ and $W_k$
theories, it is convenient to parametrize the $S^{i}$ (self-dual)
currents and the remaining anti self-dual ones ${\tilde S}^i$
\be
{\tilde S}^{i} = \frac{1}{2}\left(-\psi_{0}\psi^{i}+
\frac{1}{2}\epsilon^{ijl}\psi_{j}\psi_{l}\right)\ .
\ee
in terms of two free bosons, $H^{+}$ and $H^{-}$,  both
compactified on a circle with radius $R_{H^{+}}=R_{H^{-}}=1$
(the self-dual $SU(2)$ extended symmetry point).

In both cases, the four $N=4$ supercurrents $G$, $G^{\dag}$, $\bar G$
and $\bar G^{\dag}$ are given as \cite{k}

\be
{G} = -\left( \Pi^{\dag}e^{-\frac{i}{\sqrt{2}}H^{-}} + P^{\dag}e^{+
\frac{i}{\sqrt{2}} H^{-}} \right) e^{+ \frac{i}{ \sqrt{2}}H^{+}}
\label{compl1}\ee
\be
{\bar {G}} = \left( \Pi \;\; e^{+\frac{i}{\sqrt{2}}H^{-}} - P\;\;
e^{-\frac{i}{ \sqrt{2}} H^{-}} \right) e^{+ \frac{i}{\sqrt{2}}H^{+}}
\label{compl2}
\ee
where $P$, $P^{\dag}$ and $\Pi$, $\Pi^{\dag}$ are the four coordinate
currents.
In $\RR^4$ they are
\be
\Pi = \partial X_0+i\partial X_3 \, , \;\;\;\;\;  \Pi^{\dag}=
-\partial X_0 +i\partial X_3 \,
\ee
\be
P= \partial X_1 +i\partial X_2 \, , \;\;\;\;\;  P^{\dag}= -\partial
X_1 +i\partial X_2 \,
\label{complbF}
\ee
In the $W_k$ case,  $P$, $P^{\dag}$ and $\Pi$, $\Pi^{\dag}$ get
modified due to the torsion and non-trivial dilaton. They can be
constructed  in terms of the $SU(2)_k\otimes \RR_Q$ (anti-hermitian)
currents  $J_i$ $i=1,2,3$,  $J_0=\partial x^{0}$  and $H^{-}$,

$$
\Pi=J_0 + iQ(J_3+\sqrt{2}\partial H^-), \,\,\,\,\,
\Pi^{\dag}=-J_0+ iQ(J_3+\sqrt{2}\partial H^-)\ ,
$$
\be
P=Q(J_1+iJ_2),
\,\,\,\,\,\,\,\,\,\,\,\,\,\,\,\,\,\,\,\,\,\,\,\,\,\,\,\,\,\,\,\,\
P^{\dag}=Q(-J_1+iJ_2)~~~~~~~~~~~~~~
\label{Pcompl}
\ee

The $H^-$ modifications are due to the non-trivial background fields
of the $W_{k}$ space.
In terms of the bosonized fermions, both the standard fermionic
torsion terms $\pm Q\psi_i\psi_j\psi_l$ and the fermionic background
charge terms ($\pm Q\partial\psi_i$) are combined in such a way that
only  the anti-self dual  combination, $\partial H^-$, modifies the
coordinate currents. The other important observation is the $H^{+}$
part in the supercurrents is factorized. The $H^{+}$ factorization
property is universal in  all $N=4$ superconformal ${\hat{c}}=4$
superconformal theories \cite{k}.

It is evident from  (\ref{compl1}), (\ref{compl2}) and (\ref{Pcompl})
 that the supercurrents
(${G},G^{\dag}$) and (${\bar {G}},{\bar{G}}^{\dag}$) form two
doublets under $SU(2)_{H^{+}}$. On the other hand,
${G}$ and ${\tilde {G}}$ do not transform covariantly under the
action of $SU(2)_{H^{-}}$. They are odd, however, under a $\Z_2$
transformation, defined by $(-)^{2{\tilde S}}$, which is the parity
operator associated to the $SU(2)_{H^{-}}$ spin ${\tilde S}$ (integer
spin representations are even, while half-integer representations are
odd).

In $W_k$ we can define a global  $SU(2)_{k+1}$ charge  as the
diagonal combination of $SU(2)_k$ and $SU(2)_{H^-}$:
\begin{equation}
{\cal N}_i = J_i + {\tilde S}_i\ .
\label{ninumb}
\end{equation}

(${G},G^{\dag}$) and (${\bar {G}},{\bar {G}}^{\dag}$) form
two doublets under this $SU(2)_{k+1}$. Moreover $G$ and ${\bar {G}}$
have $({\cal N}_3, S_3)$ charges equal to $(-1/2,1/2)$ and
$(1/2,1/2)$, respectively. The  global charge ${\cal N}_3$ in $W_k$
plays the role of the helicity operator $N_h$ of flat space
\be
N_h=N_p+{\tilde S}_3
\ee
where $N_p$ is the bosonic oscillator number which counts the number
of
the $P$-oscillators minus the number of  $P^{\dag}$ ones.

The $N_h$, ${\cal N}_3$ charge, the $(-)^{2{\tilde S}}$ parity, as
well as
the $SU(2)_{H^{+}}$ spin $S$ play an important role in the
definition of the induced generalized GSO projections of the unitary
$N=4$ characters.

We would now like to show  that when flat Euclidean  four-space is
replaced by $W_k$, the maximal supersymmetry is reduced by a factor
of two. In order to avoid heavy notation for the vertex operators
which include the reparametrization ghosts, it is convenient to start
our discussion with a six dimensional theory $\RR^2\otimes
\RR^4\otimes K$ and compare it with  $\RR^2\otimes W_k\otimes K$.
In the type-II case,   both the flat  and curved constructions have
their degrees of
freedom arranged in three superconformal theories as:
\begin{equation}
\{ \hat c\} = 10 = \{ \hat c = 2 \} + \{ \hat c = 4 \}_1 + \{ \hat c
= 4\}_2~.
\label{ceq}
\end{equation}
The $\hat c = 2$ system is saturated by two free superfields. The
remaining eight supercoordinates appear
in  groups of four in $\{\hat c = 4\}_1$ and $\{\hat c = 4\}_2$. Both
$\{\hat c = 4\}_A$ systems exhibit $N = 4$ superconformal symmetry of
the Ademollo et al. type \cite{ademolo}.

The advantage of the six dimensional space in which two
super-coordinates are flat is that we can use the light-cone picture
in which the two-dimensional subspace $\RR^2$ is flat (non-compact
with
Lorentzian signature) and the eight transverse coordinates are
described by
the $\{\hat c = 4\}_1$ and $\{\hat c = 4\}_2$ theories. In this
picture,  the supersymmetry generators are constructed by
analytic (or antianalytic) dimension-one currents, whose transverse
part
is a spin-field of dimension 1/2, constructed in terms of the $H^+_A$
and $H^-_A$ bosonized fermions (of the two ${\hat c}_A=4$,
$A=1,2$ theories). In the toroidal case, there are four such
spin-fields,
\begin{eqnarray}
\Theta_{\pm} &=& e^{\frac{i}{\sqrt 2}(H^+_1 \pm H^+_2)}\nonumber \\
{\rm and}~~~~\tilde{\Theta}_{\pm} &=& e^{\frac{i}{\sqrt 2}(H^-_1 \pm
H^-_2)}\ .
\label{susy}
\end{eqnarray}
which
are even under the GSO parity:
\begin{equation}
e^{2i\pi (S_1 +S_2)}\ ,
\label{GSO}
\end{equation}
where $S_A$ are the two $SU(2)_{H^+_A}$ level-one $N=4$ spins.

In the case of  $\RR^2\otimes W_k\otimes K$ , only
the two supersymmetry generators based on the operators
$\Theta_{\pm}$, which are constructed with $H^+$'s, are
BRS-invariant. Indeed, the other two operators
$\tilde{\Theta}_{\pm}$
are not physical, due to the presence of the $\partial H^-$
modification, related to the torsion and/or background charge, in the
supercurrent expressions.

The global existence of the (chiral) $N=4$ superconformal algebra
implies in both constructions, a universal GSO projection that
generalizes \cite{worm} the one of the $N=2$ algebra
\cite{lls,gepner4d},
and which is responsible for the
existence of space-time supersymmetry. This projection restricts the
physical spectrum to being odd under the total $H^+_A$ parity
(\ref{GSO}). Thus, the supersymmetry generators based on
$\Theta_{\pm}$, which are even under (\ref{GSO}), when acting on
physical states, create physical states with the same mass but with
different statistics. The GSO projection restricts
the (level-one) character combinations associated with the two
$SU(2)_{H^+}$'s to appear in the form:
\begin{equation}
\frac{1}{2} (1-(-)^{l_1+l_2})
\chi^{H^+_1}_{l_1}\chi^{H^+_2}_{l_2} =
\chi^{H^+_1}_{l_1} \chi^{H^+_2}_{1-l_1}\delta_{l_2,1-l_1}\ ,
\label{charplus}
\end{equation}
with $l_A=2S_A$ taking values 0 or 1, corresponding to the two
possible characters, (spin-$0$ and spin-$1/2$)  of the $SU(2)_{1}$
affine algebra.

The basic rules in both  constructions are similar to those of the
orbifold construction \cite{orbifold}, the free 2-d
fermionic constructions \cite{ferm},  and the Gepner construction
\cite{gepner4d}. There, one combines in a modular invariant way the
world-sheet degrees of freedom consistently with unitarity and
spin-statistics of  the string spectrum. In both cases, the 2-d
fermions are free and their characters are given in terms of
$\vartheta$-functions. The 6-d Lorentz invariance in the flat case
and the existence of $SU(2)_k$ currents in  the curved case imply the
same boundary conditions for  the super-reparametrization ghosts and
the six of the worldsheet fermions. In the flat case there is no
obstruction to choose  the remaining four fermions with the same
boundary conditions and obtain the well known partition function with
maximal space-time supersymmetry:

\begin{eqnarray}
Z_F = {1\over \im^2|\eta|^8}~{1\over 4}
\sum_{\alpha,\beta\atop {\bar\alpha},{\bar\beta}}~
(-)^{\alpha+\beta+{\bar\alpha}+{\bar\beta}}
\frac{\vartheta^{2}[^{\alpha}_{\beta}]}{\eta^2}
\frac{\vartheta^{2}[^{\alpha}_{\beta}]}{\eta^2}~
\frac{{\bar\vartheta}^{2}[^{\bar\alpha}_{\bar\beta}]}{{\bar\eta}^2}
\frac{{\bar\vartheta}^{2}
    [^{{\bar\alpha}}_{{\bar\beta}}]}{{\bar\eta}^2}~Z_4[^0_0],
\label{pF}
\end{eqnarray}
where the $Z_4[^0_0]$ contribution is that of four compactified
coordinates:
\be
Z_{4}[^0_0]={\Gamma(4,4)\over |\eta|^8}\label{444}
\ee
and $\Gamma(4,4)$ stands for the usual lattice sum.  $\alpha$,
$\beta$ and ${\bar\alpha}$, ${\bar\beta}$ denote the left- and
right-moving  spin structures.  The spin-statistic factors
$(-)^{\alpha +\beta}$ and
$(-)^{\bar{\alpha} +\bar{\beta}}$ come from the contribution of the
(left- and right-moving) $\RR^2$ world-sheet fermions and the
(left- and right-moving) $({\bf\beta}, {\bf\gamma})$-ghosts.
The Neveu-Schwarz ($NS$, $\overline{NS}$) sectors correspond to
$\alpha, \bar{\alpha}=0$ and the Ramond ($R,\bar{R}$) sectors
correspond to $\alpha, \bar{\alpha}=1$. For later convenience we
decompose the $O(4)$ level-one characters, which are written in terms
of
$\vartheta$-functions, in terms of the $SU(2)_{H^{+}_{1}}\otimes
SU(2)_{H^{-}_{1}}$ characters using the identity:
\begin{equation}
\frac{\th ^{2} [^{\alpha}_{\beta}]}{ \eta ^{2}(\tau)}=
\sum_{l=0}^{1} (-)^{\beta l} \chi_{l}^{H^{+}_1}
\chi_{l+\alpha(1-2l)}^{H^{-}_1}\ ,
\label{theeta}
\end{equation}
and similarly for the right-movers.

Using the decomposition above, we can write the flat partition
function in terms of $SU(2)_{H^{+}_1}$, $SU(2)_{H^{-}_2}$ ,
$SU(2)_{H^{+}_2}$ and
 $SU(2)_{ H^{-}_2 }$ characters as

\begin{eqnarray}
Z_F = {1\over \im^2|\eta|^8}
{}~\sum_{\alpha,{\bar\alpha}}~
(-)^{\alpha+{\bar\alpha}}~
\chi_{l}^{H^{+}_1} \chi_{l+\alpha}^{H^{-}_1}~
\chi_{1+l}^{H^{+}_2} \chi_{1+l+\alpha}^{H^{-}_2}~
{\bar\chi}_{\bar l}^{H^{+}_1}
{\bar\chi}_{{\bar l}+{\bar\alpha}}^{H^{-}_1}~
{\bar\chi}_{1+{\bar l}}^{H^{+}_2}
{\bar\chi}_{1+{\bar l}+{\bar\alpha}}^{H^{-}_2}
{}~Z_4[^0_0],
\label{pFchar}
\end{eqnarray}

In going   from (\ref{pF}) to (\ref{pFchar}), the $\beta$ and
${\bar\beta}$ summations give rise to the universal (left- and
right-moving) GSO projections among the $SU(2)_{H^{+}_1}$ and
$SU(2)_{H^{+}_2}$ spins, $2 S_1 + 2{ S}_2$=odd, as well as among
those of $SU(2)_{H^{-}_1}$, $SU(2)_{H^{-}_2}$, $2\tilde S_{1}+2\tilde
S_{2}=$ odd. These projections
imply the existence of
maximal space-time
supersymmetry. The phase $(-)^{\alpha+{\bar\alpha}}$ guarantees the
spin-statistics connection; it equals $+1$ for space-time bosons and
$-1$ for space-time fermions.

Replacing $\RR^4$ flat space with  $W_k$, some modifications are
necessary. Namely, we must combine the $\RR_Q$ Liouville-like
characters and  the $SU(2)_k$ ones ($\chi _{L}~,~L=0,1,2,\cdots, k$)
with those of  the remaining  2-d bosons and fermions, in a way
consistent with unitarity and modular invariance.

The $\RR_Q$ Liouville-like characters can be classified in two
categories. Those that correspond to the continuous representations
generated by the  operators:
\begin{equation}
e^{\beta X_L}\ ;\quad \beta=-\frac{1}{2}Q +ip\ ,
\label{contchar}
\end{equation}
having positive conformal weights
$\Delta_p=\frac{Q^2}{8}+\frac{p^2}{2}$.
The fixed imaginary part in the momentum $iQ/2$ of the plane waves,
is due to the non-trivial dilaton motion. The second category of
Liouville characters \cite{sbk} corresponds to lowest-weight
operators
(\ref{contchar}) with $\beta=Q\tilde{\beta}$ real, leading to
negative conformal dimensions
$-\frac{1}{2}\tilde{\beta}(\tilde{\beta}+1)Q^2=
-\frac{\tilde{\beta}(\tilde{\beta}+1)}{k+2}$. Both categories of
Liouville representations give rise to unitary representations of
the $N=4$, ${\hat c}=4$  $W_k$ theory, once they are combined
with
the remaining degrees of freedom. The continuous representations
(\ref{contchar}) form long (massive) representations \cite{ademolo}
of $N=4$ with
conformal weights larger than the $N=4$, $SU(2)$ spin, $\Delta>S$. On
the
other hand, the second category contains short representations of
$N=4$ \cite{ademolo}
($\Delta=S$), while $\beta$ can take only a finite number of values,
$-(k+2)/2\le\tilde{\beta}\le k/2$. In fact, their locality with
respect to the
$N=4$  operators implies \cite{worm}:
\begin{eqnarray}
S &=& \frac{1}{2},\quad \tilde{S}=\frac{1}{2}:\quad \tilde{\beta} =
-(j+1) \nonumber \\
S &=& 0,\quad \tilde{S}=0:\quad \tilde{\beta} = j\ .
\label{discrchar}
\end{eqnarray}

In both cases of (\ref{discrchar}), the conformal weight $\Delta=S$
is
independent of $SU(2)_k$ and $SU(2)_{H^-}$ spins, due to the
cancellation between the Liouville and $SU(2)_k$ contributions. The
states associated to the short representations of $N=4$ do not have
the interpretation of propagating states, but they describe a
discrete set
of localized states. They are similar to the discrete states found in
the
$c=1$ matter system coupled to the Liouville field \cite{lz} and  the
two-dimensional $SL(2,R)/O(1,1)$ coset model \cite{sl2}.
Although they
play a crucial role in scattering amplitudes, they do not correspond
to asymptotic
states and they  do not  contribute to the partition function.
Indeed,
in our case they are not only discrete but also
finite in number. They obviously are of zero measure  compared to the
contribution of the continuous (propagating) representations.

The presence of discrete
representations with $\beta$ positive are necessary to define
correlation functions. In fact, the balance of the background
charge for an $N$-point amplitude at genus $g$ implies the relation
\cite{worm}
\begin{equation}
N+2(g-1)+2\sum_I\tilde{\beta}_I=0\ ,
\label{screen}
\end{equation}
where the sum is extended over the vertices of the discrete
representation states. Thus, these vertices define an appropriate set
of screening operators, necessary to define amplitudes in the
presence
of non-vanishing background charge. In our case, the screening
procedure has an interesting physical interpretation similar to the
scattering of asymptotic propagating states (continuous
representations) in the presence of non-propagating bound states
(discrete representations). The screening operation then describes
the
possible angular momentum (of SU(2)) excitations of the bound states.
Below, we
restrict ourselves to the one-loop partition function, where the
discrete representations are not necessary (see eq.(\ref{screen})
with
$g=1$ and $N=0$).

It is convenient to define appropriate character combinations of
$SU(2)_k$, which transform covariantly under modular transformations
\cite{worm}:
\begin{equation}
Z_{so(3)}[^{\alpha}_{\beta}]
=Z_{so(3)}[^{\a+2}_{\beta}]=Z_{so(3)}[^{\a}_{\beta+2}]=
e^{-i\pi\a\beta k/2}\sum_{l=0}^k e^{i\pi\beta l} \chi_l
\bar{\chi}_{(1-2\a)l+\alpha k}
\label{zab}
\end{equation}
where $\alpha$, $\beta$ can be either 0 or 1.
The $SU(2)_{k}$ characters are given by the familiar expressions
\cite{kp},
\be
\chi_{l}(\tau)={\th_{l+1,k+2}(\tau,v)-\th_{-l-1,k+2}(\tau,v)\over
\th_{1,2}(\tau,v)-\th_{-1,2}(\tau,v)}|_{v=0}
\ee
where
\be
\th_{m,k}(\tau,v)\equiv \sum_{n\in \Z}\exp\left[2\pi i
k\left(n+{m\over 2k}\right)^2\tau-2\pi i k\left(n+{m\over
2k}\right)v\right]
\ee
are the level-k $\th$-functions.
The projection induced in (\ref{zab}) will project $SU(2)\to SO(3)$.
This can be done consistently when $k$ is an even integer, which we
assume from now on.

Under modular
transformations, $Z_{so(3)}[^{\alpha}_{\beta}]$ transforms as:
\begin{eqnarray}
\tau\rightarrow\tau +1~~~:~~~
&Z_{so(3)}[^{\alpha}_{\beta}]& \longrightarrow
Z_{so(3)}[^{~\alpha}_{\beta+\alpha}]
\nonumber \\
\tau\rightarrow{-1/\tau}~~~:~~~
&Z_{so(3)}[^{\alpha}_{\beta}]& \longrightarrow
Z_{so(3)}[^{\beta}_{\alpha}]\ .
\label{zabtransf}
\end{eqnarray}

The partition function must satisfy two basic constraints emerging
from the $N=4$ algebra. The first is associated to the two spectral
flows of the $N=4$ algebra which impose the universal GSO projection
$2(S_1+S_2)$=odd among the $H^+$ spins. The second
constraint is associated to the reduction of space-time
supersymmetries by a factor of 2. It imposes a second projection
which eliminates half of the lowest-lying states constructed from the
$H^-_1$ field, which are not local with respect to the
$N=4$ generators. These unphysical states should be eliminated from
the spectrum by an additional GSO projection involving the two
$H^{-}_i$ spins
$\tilde S_i$ as well as the spin of $SU(2)_k$.

For $k$ even, there is a $\Z_2$ automorphism of $SU(2)_k$ which
leaves
invariant the currents but acts non-trivially on the
odd
spin representations. This allows to correlate the $SU(2)_{H_1^-}$,
$SU(2)_{H_1^-}$
and
$SU(2)_k$ spins in a way which projects out of the spectrum the
unphysical states. This $\Z_2$ must act simultaneously   on the four
toroidal compactified coordinates in order to guarantee the global
existence of the $N=1$ supercurrent. The modular-invariant partition
function then is:
\be
Z_W = {1\over \im^{1/2}|\eta|^2}~{1\over 8}
\sum_{\alpha,\beta,{\bar\alpha}\atop {\bar\beta},\gamma,\delta}
(-)^{(\alpha+{\bar\alpha})(1+\delta)+\beta+{\bar\beta}}
\frac{\th^{2}[^{\alpha}_{\beta}]}{\eta^2}
\frac{\th^{2}[^{\alpha+\gamma}_{\beta+\delta}]}{\eta^2}~
\frac{{\bar\th}^{2}[^{\bar\alpha}_{\bar\beta}]}{{\bar\eta}^2}
\frac{{\bar\th}^{2}
[^{{\bar\alpha}+\gamma}_{{\bar\beta}+\delta}]}{{\bar\eta}^2}
{Z_{so(3)}[^{\gamma}_{\delta}]\over V}~ Z_4[^\gamma _\delta]
\label{pW}
\ee
where $Z_4[^\gamma _\delta ]$ denotes the $T^{(4)}/\Z_2$ orbifold
twisted characters. $Z_{4}[^0_0]$ is given in (\ref{444}) while for
$(h,g)\not= (0,0)$
we have
\be
Z_{4}[^h_g]={|\eta|^4\over |\th[^{1+h}_{1+g}]\th[^{1-h}_{1-g}]|^2}
\ee
We have also divided by the (quantum) volume of $S^3$
\be
V={(k+2)^{3/2}\over 8\pi}\;\;\;.\label{volume}
\ee

We may rewrite the partition function above in terms of  various
$SU(2)$ characters so that the induced GSO projections are more
transparent.
$$
Z_W = {1\over \im^{1/2}|\eta|^2}
{}~\sum_{\alpha,{\bar\alpha},\gamma,l,{\bar l}=0}^{1}
(-)^{\alpha+{\bar\alpha}}~
\chi_{l}^{H^{+}_1} \chi_{l+\alpha}^{H^{-}_1}~
\chi_{l+1}^{H^{+}_2} \chi_{l+1+\alpha+\gamma}^{H^{-}_2}~
{\bar\chi}_{\bar l}^{H^{+}_1}
{\bar\chi}_{{\bar l}+{\bar\alpha}}^{H^{-}_1}~
{\bar\chi}_{1+{\bar l}}^{H^{+}_2}
{\bar\chi}_{1+{\bar l}+{\bar\alpha}+\gamma}^{H^{-}_2}\times
$$
\be
\times~{1\over V}\sum_{L=0}^{k}\sum_{\delta=0}^{1} {1\over 2}
(-)^{\delta[\alpha+l+{\bar\alpha}+{\bar l}+{k\over 2}\gamma+L]}~
\chi_L \bar{\chi}_{(1-2\gamma)L+\gamma k}~Z_4[^{\gamma}_{\delta}]\ ,
\label{pWchar}
\ee

As in the flat case, the $\beta$ and
${\bar\beta}$ summations give rise to the universal (left- and
right-moving) GSO projections $2(S_1+S_2)$=odd, which imply the
existence of
space-time
supersymmetry. The summation over $\delta$ however gives rise to an
additional projection, which correlates the $SU(2)_{H^-_2}$ (left and
right) spin together with the spin of $SU(2)_k$ and $T^{(4)}/\Z_2$
twisted bosonic oscillator numbers. This projection reduces
the number of space-time supersymmetries by a factor of two.

In the $\gamma=0$ sector (untwisted sector), the lower-lying states
have (left and
right)
mass-squared $Q^2/8$ and $L=0$. This is due to the non-trivial
dilaton for the bosons,
and to the non-trivial torsion for the fermions.

The contribution of 2-d fermions in the partition function
of the $\gamma=0$ sector is identical to the fermionic part of the
partition function of the ten-dimensional type II superstring with
an additional projection acting on $SU(2)_k$ spins :
\begin{equation}
Z_W^{\gamma =0} = {1\over \im^{1/2}|\eta|^2}
{1\over 4} |\th_3^4-\th_4^4-\th_2^4|^2
\sum_{L=even}^{k}{ |\chi_L|^2\over V}~Z_4[^0_0]
\label{pfu}
\end{equation}

As was stressed before, the extra  $\Z_2$ projection  is dictated
from the $N=4$ superconformal algebra, in order to eliminate the
unphysical states from the
untwisted sector. Modular invariance implies
the presence of a twisted sector $(\gamma=1)$, which contains states
with (left and right) mass-squared always larger than $(k-2)/16$.
In the large $k$ limit the twisted states become super-heavy

We can now return to our initial problem and examine the Euclidean
$W_k\otimes K$ theory with maximal space-time supersymmetry. The
latter can be obtained by a $T^{2}$ torus compactification from the
Euclidean version
of the six dimensional construction described above. Observe that it
is necessary to  act non-trivially in the internal theory $K$, since
the $\Z_2$ in question has to act on the four out of the six
compactified coordinates.
The resulting partition function is:
\be
Z^{4d}_W = {\im^{1/2}|\eta|^2\over 8}
\sum_{\alpha,\beta,{\bar\alpha}\atop {\bar\beta},\gamma,\delta}
(-)^{(\alpha+{\bar\alpha})(1+\delta)+\beta+{\bar\beta}}
\frac{\th^{2}[^{\alpha}_{\beta}]}{\eta^2}
\frac{\th^{2}[^{\alpha+\gamma}_{\beta+\delta}]}{\eta^2}
\frac{{\bar\th}^{2}[^{\bar\alpha}_{\bar\beta}]}{{\bar\eta}^2}
\frac{{\bar\th}^{2}
[^{{\bar\alpha}+\gamma}_{{\bar\beta}+\delta}]}{{\bar\eta}^2}
{Z_{so(3)}[^{\gamma}_{\delta}]\over V}Z_2[^0_0]Z_4[^\gamma _\delta]\
,
\label{pW4}
\ee
where $Z_2[^0_0]$ is the contribution of the $T^{(2)}$
compactification,
on a $(2,2)$ Lorentzian lattice: $Z_2[^0_0]$=$ \Gamma
(2,2)/|\eta|^4$.

In the heterotic case, a modular-invariant partition function for $k$
even can be easily obtained using the heterotic map \cite{lls},
\cite{gepner4d}. It consists of
replacing in (\ref{pW4}) the $O(4)$ characters associated to the
right-moving fermionic coordinates ${\bar\psi}^{\mu}$, with the
characters of either $O(12)\otimes E_8$:
\begin{equation}
(-)^{{\bar\alpha}+{\bar\beta}}
\frac{{\bar\th}^{2}[^{\bar\alpha}_{\bar\beta}]}{{\bar\eta}^2}
\rightarrow
\frac{{\bar\th}^{6}[^{\bar\alpha}_{\bar\beta}]}{{\bar\eta}^6}
{1\over
2}\sum_{\gamma\delta=0}^{1}{\bar\th^8[^{\gamma}_{\delta}]\over
\bar\eta^8}
\label{pfheta}
\end{equation}
or $O(28)$:
\begin{equation}
(-)^{{\bar\alpha}+{\bar\beta}}
\frac{{\bar\th}^{2}[^{\bar\alpha}_{\bar\beta}]}{{\bar\eta}^2}
\rightarrow
\frac{{\bar\th}^{14}[^{\bar\alpha}_{\bar\beta}]}{{\bar\eta}^{14}}
\ .
\label{pfhetb}
\end{equation}

Other heterotic constructions can be obtain in the case where the
extra $\Z_2$
projection acts asymmetrically on the left and right degrees of
freedom.
In all these constructions the number of space time supersymmetries
in flat 4d space
($N=8$ in type II and $N=4$ in heterotic) is reduced by a factor
of two when we move in the $W_k$ space.

This reduction of space time supersymmetries due to the non trivial
mixing of
the $SU(2)_k$ characters and those of the internal space can be
avoided
in the case where the flat construction has a lower number of space
time supersymmetries. In order to see how this works, we will examine
first the case of $\Z_2$ symmetric orbifold, based on  $\RR^4\otimes
T^{(2)}\otimes T^{(4)}/\Z_2$, in which the number of supersymmetries
is $N=2$
in heterotic and $N=4$ in type II.
Contrary to the maximal supersymmetry case, here the number of
supersymmetry
is already reduced by the $\Z_2$ orbifold projection which acts non
trivially
to the two spin fields $\tilde{\Theta}_{\pm} = e^{\frac{i}{\sqrt
2}(H^-_1 \pm
H^-_2)}\ $
constructed with the $H^-_i$ bosons.
The $\Z_2$ orbifold partition function for $\RR^4\otimes
T^{(2)}\otimes T^{(4)}/\Z_2$ is:

\be
Z^{\Z_2} = {1\over \im|\eta|^4}~{1\over 8}
\sum_{\alpha,\beta,{\bar\alpha}\atop {\bar\beta},h,g}
(-)^{\alpha+\beta+{\bar\alpha}+{\bar\beta}}
\frac{\th^{2}[^{\alpha}_{\beta}]}{\eta^2}
\frac{\th^{2}[^{\alpha+h}_{\beta+g}]}{\eta^2}~
\frac{{\bar\th}^{2}[^{\bar\alpha}_{\bar\beta}]}{{\bar\eta}^2}
\frac{{\bar\th}^{2}
[^{{\bar\alpha}+h}_{{\bar\beta}+g}]}{{\bar\eta}^2}
{}~Z_2[^0_0]Z_4[^h _g]
\label{pFZ2}
\ee
The $g$-action projects out in  the  untwisted sector (h=0) the
unwanted spin fields, $\tilde{\Theta}_{\pm}=e^{\frac{i}{\sqrt
2}(H^-_1 \pm
H^-_2)}\ $ as usual.

 Replacing flat space $\RR^4$ with $W_k$ in the  orbifold model
above,
we must specify the extra $\Z_2$ action, which, as we explained in
the maximal supersymmetry example, must act non-trivially on the
$\tilde{\Theta}_{\pm}=e^{\frac{i}{\sqrt 2}(H^-_1 \pm
H^-_2)}\ $ spin fields as well as on the $SU(2)_k$ characters. This
action must be in agreement with modular invariance and unitarity.
The resulting partition function is:

\be
Z_{W}^{\Z_2} = {\im^{1/2}|\eta|^2\over 16}
\sum_{\alpha,\beta,{\bar\alpha},{\bar\beta}\atop\gamma,\delta,h,g}
(-)^{\alpha+\beta+{\bar\alpha}+{\bar\beta}}
\frac{\th^{2}[^{\alpha+\gamma}_{\beta+\delta}]}{\eta^2}
\frac{\th^{2}[^{\alpha+h}_{\beta+g}]}{\eta^2}
\frac{{\bar\th}^{2}[^{\bar\alpha+\gamma}_{\bar\beta+\delta}]}
{{\bar\eta}^2}
\frac{{\bar\th}^{2}
[^{{\bar\alpha}+h}_{{\bar\beta}+g}]}{{\bar\eta}^2}
{Z_{so(3)}[^{\gamma}_{\delta}]\over V} Z_2[^0_0]Z_4[^{h+\gamma}
_{g+\delta }]
\label{pFZ3}
\ee

Redefining the parameters $\alpha\rightarrow\alpha-\gamma$,
$\beta\rightarrow\beta-\delta$, $h\rightarrow h+\gamma$ $g\rightarrow
g+\delta$,
the partition function above takes the following factorized form:
\be
Z_{W}^{\Z_2} = \im^{1/2}|\eta|^2~{1\over 2}\sum_{\gamma,\delta}
{Z_{so(3)}[^{\gamma}_{\delta}]\over V}~{1\over 8}
\sum_{\alpha,\beta,{\bar\alpha}\atop {\bar\beta},h,g}
(-)^{\alpha+\beta+{\bar\alpha}+{\bar\beta} }
\frac{\th^{2}[^{\alpha}_{\beta}]}{\eta^2}
\frac{\th^{2}[^{\alpha+h}_{\beta+g}]}{\eta^2}
\frac{{\bar\th}^{2}[^{\bar\alpha}_{\bar\beta}]}{{\bar\eta}^2}
\frac{{\bar\th}^{2}
[^{{\bar\alpha}+h}_{{\bar\beta}+g}]}{{\bar\eta}^2}
Z_2[^0_0]Z_4[^{h} _{g}]
\label{pFZ2fact}
\ee
Using the heterotic map (\ref{pfheta}) or (\ref{pfhetb})) we obtain
heterotic constructions
in curved space with $N=2$ spacetime supersymmetric spectrum. Since
the $SO(3)_{k/2}$ contribution factorizes, the number of space-time
supersymmetries remains the same as in  flat space.

The factorization property above, is not a special property of the
$\Z_2$ symmetric orbifold but is generic for all $\RR^4\otimes K$
models provided the internal space $K$ is not a theory that produces
maximal supersymmetry.
Indeed, in cases where the internal $K$ theory is non-trivial,  the
spin fields which are non-BRST invariant in $W_k\otimes K$ are
already absent in the flat space ground state. The validity of this
statement can be proven  in all orbifold constructions.

The reason of the non-factorization in the case where $K$ is toroidal
is due to the reduction by a factor of two  of the covariantly
constant spinors in the $W_k$ background indicating that spacetime
supersymmetry cannot be maximal. When the internal space is not
trivial then the number of space time supersymmetries is already
reduced in flat space and thus, further reduction is not necessary.
Observe however, that the odd spin representations  of the $SU(2)_k$
($2j$=odd) are absent, since they are projected out by the $\Z_2$
projection discussed above. Therefore, the correct target space is
the $\Z_2$ orbifold of $SU(2)_k$, namely
that of $SO(3)_{k/2}$.

Thanks to the factorization property described above in the
non-maximal supersymmetric case, we can always construct the curved
space-time partition function, $Z^W(\tau,{\bar \tau})$ in terms of
that of flat space $Z_{0}(\tau,{\bar \tau})$,
\be
Z^W(\tau,{\bar \tau})=
\im^{3/2}|\eta(\tau)|^6~{\Gamma(SO(3)_{k/2})\over V}Z_{0}(\tau,{\bar
\tau})\label{zero}
\ee
where $\Gamma(SO(3)_{k/2})$ is the partition function of the $SO(3)$
WZW model at level $k/2$:
\be
\Gamma(SO(3)_{k/2})={1\over
2}\sum_{\gamma,\delta=0}^{1}Z_{so(3)}[^{\gamma}_{\delta}]
\ee

\section{The $\sigma$-model description of magnetic and
gravitational backgrounds}\setcounter{equation}{0}

The starting 4-d spacetime (we will use Euclidean signature here) is
described
by the $SO(3)_{k/2}\times \RR_{Q}$ CFT.
The heterotic $\sigma$-model that describes this space is\footnote{
In most formulae we set $\alpha'=1$ unless stated otherwise.}

\be
S_{4d}={k\over 4}{\bf I}_{SO(3)}(\a,\b,\g)
+{1\over 2\pi}\int d^2z \left[\p x^{0}\pb
x^{0}+\psi^{0}\pb\psi^{0}+\sum_{a=1}^{3}\psi^{a}\pb\psi^{a}\right]+
{Q\over 4\pi}\int \sqrt{g}R^{(2)}x^{0}
\ee
while the SU(2) action can be written in Euler angles as
\be
{\bf I}_{SO(3)}(\a,\b,\g)={1\over 2\pi}\int d^2
z\left[\p\a\pb\a+\p\b\pb\b
+\p\g\pb\g+2\cos\b\p\a\pb\g\right]
\ee
with $\b\in[0,\pi]$, $\a,\g\in[0,2\pi]$ and $k$ is a positive even
integer.
In the type II case we have to add also the right moving fermions
$\bar \psi^{i}$, $1\leq i\leq 4$. The fermions are free (this is a
property
valid for all supersymmetric WZW models).

Comparing with the general (bosonic) $\sigma$-model
\be
S={1\over 2\pi}\int d^2 z (G_{\mu\nu}+B_{\mu\nu})\p x^{\mu}\pb
x^{\nu}+
{1\over 4\pi}\int \sqrt{g}R^{(2)}\Phi(x)
\ee
we can identify the non-zero background fields as
\be
G_{00}=1\;\;,\;\; G_{\a\a}=G_{\b\b}=G_{\g\g}={k\over 4}
\label{3s1}
\ee
\be
G_{\a\g}={k\over 4}\cos\b\;\;\;,\;\;\;B_{\a\g}={k\over 4}\cos\b
\label{3s2}\ee
\be
\Phi=Qx^{0}={x^{0}\over \sqrt{k+2}}\label{dil}
\ee
where the relation between $Q$ and $k$ is required from the
requirement that
the (heterotic) central charge should be $(6,4)$, in which case we
have (4,0)
superconformal invariance, \cite{k}.

The perturbation that turns on a chromo-magnetic field in the $\mu=3$
direction
is proportional to $(J^{3}+\psi^1\psi^2)\bar J$ where $\bar J$ is a
right moving current belonging to the Cartan subalgebra of the
heterotic gauge group.
It is normalized so that $\langle\bar J(1)\bar J(0)\rangle=k_{g}/2$.
Since
\be
J^{3}=k(\p\g+\cos\b\p\a)\;\;\;,\;\;\;J^{3}=k(\pb\a+\cos\b\pb\g)
\ee
this perturbation changes the $\sigma$-model action in the following
way:
\be
\delta S_{4d}={\sqrt{kk_{g}}H\over 2\pi}\int d^2
z(\p\g+\cos\b\p\a)\bar J
\label{gauge}
\ee

In the type II case $\bar J$ is a bosonic current (it has a left
moving partner) and we can easily show that the $\sigma$-model with
action
$S_{4d}+\delta S_{4d}$ is conformally invariant to all orders in
$\alpha'$.
This can be seen by writing $\bar J=\pb \phi$ and noticing that

\be
{k\over 4}{\bf I}_{SO(3)}(\a,\b,\g)+\delta S_{4d}+{k_{g}\over
4\pi}\int d^2
z~\p\phi\pb\phi={k\over 4}{\bf
I}_{SO(3)}\left(\a,\b,\g+2\sqrt{k_{g}\over
k}H\phi\right)+
\label{wzw}\ee
$$+{k_{g}(1-2H^2)\over 4\pi}\int d^2
z~\p\phi\pb\phi
$$

It is already obvious from  (\ref{wzw}) that something special
happens at
$H^2=1/2$. In fact in the toroidal case that would correspond to a
boundary
of moduli space. It is the limit Im$U\to \infty$ in the case of a
$(2,2)$ lattice.
Here the interpretation would be of a maximum magnetic field.
We will see more signals of this later on.

Reading the spacetime backgrounds from (\ref{gauge}) is not entirely
trivial but straightforward.
In type II case (which corresponds to standard Kalutza-Klein
reduction) the correct metric  has an $A_{\mu}A_{\nu}$ term
subtracted \cite{dpn}.
In the heterotic case there is a similar subtraction but the reason
is different. It has to do with the anomaly in the holomorphic
factorization
of a boson (see for example \cite{p}).

The background fields have to be solutions (in leading order in $\a
'$) to the following equations of motion \cite{eff}:

\be
\delta c={3\over 2}\left[4(\n\Phi)^2-{10\over 3}\square\Phi-{2\over
3}R+{1\over 12g^2}F^{a}_{\mu\nu}F^{a,\mu\nu}\right]=0
\label{1}\ee

\be
R_{\mu\nu}-{1\over 4}H^2_{\mu\nu}-{1\over
2g^2}F^a_{\mu\rho}{F^{a}_{\nu}}^{\rho}+2\n_{\mu}\n_{\nu}
\Phi=0\label{2}
\ee

\be
\n^{\mu}\left[e^{-2\Phi}H_{\mu\nu\rho}\right]=0\label{3}
\ee

\be
\n^{\nu}\left[e^{-2\Phi}F^{a}_{\mu\nu}\right] -{1\over
2}F^{a,\nu\rho}
H_{\mu\nu\rho}e^{-2\Phi}\label{4}
\ee
which stem from the variation of the effective action,

\be
S=\int d^{4}x\sqrt{G}e^{-2\Phi}\left[R+4(\n\Phi)^2-{1\over
12}H^2-{1\over 4g^2}F^a_{\mu\nu}F^{a,\mu\nu}+{\delta c\over
3}\right]\label{5}
\ee
where we have displayed a  gauge field $A^{a}_{\mu}$, (abelian or
non-abelian) and set $g_{\rm string}=1$. The gauge coupling
is $g^2=2/k_{g}$ due to the normalization of the currents in
(\ref{norm}),

\be
F^a_{\mu\nu}=\p_{\mu}A_{\nu}-\p_{\nu}A_{\mu}+f^{abc}
A^{b}_{\mu}A^{c}_{\nu}
\label{6}
\ee

\be
H_{\mu\nu\rho}=\p_{\mu}B_{\nu\rho}-{1\over
2g^2}\left[A^a_{\mu}F^a_{\nu\rho}-{1\over
3}f^{abc}A^{a}_{\mu}A^{b}_{\nu}A^{c}_{\rho}\right]+{\rm
cyclic}\;\;{\rm permutations}\label{7}
\ee
and $f^{abc}$ are the structure constants of the gauge group.
In this paper we will restrict ourselves to abelian gauge fields (in
the cartan
of a non-abelian gauge group).

It is not difficult now to read from (\ref{gauge}) the background
fields
that satisfy the equations above.
The non-zero components are:
\be
G_{00}=1\;\;,\;\;G_{\b\b}={k\over 4}\;\;,\;\;G_{\a\g}={k\over
4}(1-2H^2)\cos\b
\label{8}\ee
\be
G_{\a\a}={k\over 4}(1-2H^2\cos^2\b)\;\;,\;\;G_{\g\g}={k\over
4}(1-2H^2)\;\;,\;\;B_{\a\g}={k\over 4}\cos\b
\label{9}\ee
\be
A_{a}=g\sqrt{k}H\cos\b\;\;\;,\;\;\;A_{\g}=g\sqrt{k}H
\label{10}\ee
and the same dilaton as in (\ref{dil}).
As shown before this background is exact to all orders in the $\a '$
expansion
with simple modification $k\to k+2$.

It is interesting to note that
\be
\sqrt{{\rm det}G}=\sqrt{1-2H^2}\left({k\over 4}\right)^{3/2}\sin\b
\ee
which indicates, as advertised earlier, that something special
happens
at $H_{max}=1/\sqrt{2}$.
At this point the curvature is regular.
In fact, this is a boundary point where the states that couple to the
magnetic field (i.e. states with non-zero $\Q+I$ and/or $e$) become
infinitely massive and decouple.
This is the same phenomenon as the degeneration of the Kh\"aler
structure on a two-torus (${\rm Im}U\to\infty$).
Thus, this point is at the boundary of the magnetic field moduli
space.
This is very interesting since it implies the existence of a maximal
magnetic
field
\be
|H|\leq H_{max}={1\over \sqrt{2}}
\ee
We should note here that the deformation of the spherical geometry by
the magnetic field is smooth for all ranges of parameters, even at
the boundary point $H=1/\sqrt{2}$.
To monitor better the back-reaction of the effective field theory
geometry
we should first write the three-sphere with the round metric
(\ref{3s1}), (\ref{3s2}), as the (Hopf)
fibration of $S^1$ as fiber and a two-sphere as base space:
\be
ds^2_{\rm 3-sphere}={k\over 4}\left[ds^2_{\rm 2-sphere}+(d\g+\cos\b
d\a)^2\right]
\label{hopf}
\ee
with
\be
ds^2_{\rm 2-sphere}=d\b^2+\sin^2\b d\a^2
\ee
The second term in (\ref{hopf}) is the metric of the $S^1$ fiber, and
its non-trivial dependence on $\a,\b$ signals the non-triviality of
the Hopf fibration.
This metric has $SO(3)\times SO(3)$ symmetry.

The metric (\ref{8}), (\ref{9}) containing the backreaction to the
non-zero magnetic field can be written as
\be
ds^2={k\over 4}\left[ds^2_{\rm 2-sphere}+(1-2H^2)(d\g+\cos\b
d\a)^2\right]
\label{back}\ee
It is obvious from (\ref{back}) the magnetic field changes the radius
of the fiber and breaks the $SO(3)\times SO(3)$ symmetry to the
diagonal $SO(3)$.
It is also obvious that at $H=1/\sqrt{2}$, the radius of the fiber
becomes
zero.
All the curvature invariants are smooth (and constant due to the
$SO(3)$ symmetry)

As mentioned in the introduction, we have another marginal
deformation
(\ref{gravi}) which turns on curvature as well as antisymmetric
tensor and dilaton.
The essential part of this perturbation is the $J^{3}\bar J^3$ part
which deforms the Cartan torus of SO(3) and the exact bosonic
$\sigma$-model
action was given in \cite{hs,gk}.
We will use this result to derive the background fields associated
with
both gauge and gravitational deformation. After some algebra we
obtain
\def\l{\lambda}
\be
G_{00}=1\;\;\;G_{\b\b}={k\over 4}\label{11}
\ee

\be
G_{\a\a}={k\over 4}{(\l^2+1)^2-(8H^2\l^2+(\l^2-1)^2)\cos^2\b\over
(\l^2+1+(\l^2-1)\cos\b)^2}
\label{14}\ee

\be
G_{\g\g}={k\over 4}{(\l^2+1)^2-8H^2\l^2-(\l^2-1)^2\cos^2\b\over
(\l^2+1+(\l^2-1)\cos\b)^2}\label{15}\ee

\be
G_{\a\g}={k\over 4}{4\l^2(1-2H^2)\cos\b+(\l^4-1)\sin^2\b\over
(\l^2+1+(\l^2-1)\cos\b)^2}\label{16}
\ee

\be
B_{\a\g}={k\over 4}{\l^2-1+(\l^2+1)\cos\b\over
(\l^2+1+(\l^2-1)\cos\b)}
\label{17}
\ee

\be
A_{a}=2g\sqrt{k}{H\l\cos\b\over (\l^2+1+(\l^2-1)\cos\b)}\label{18}
\ee

\be
A_{\g}=2g\sqrt{k}{H\l\over (\l^2+1+(\l^2-1)\cos\b)}\label{19}\ee

\be
\Phi={t\over \sqrt{k+2}}-{1\over 2}\log\left[\l+{1\over
\l}+\left(\l-{1\over \l}
\right)\cos\b\right]\label{20}
\ee
It is straightforward to verify that the fields above solve the
equations of the effective field theory.

We now have an additional modulus which governs the gravitational
perturbation,
namely $\l$ which we can take it to be a non-negative  real number.
There are however duality symmetries that act on the moduli $H,\l$.
The first is a $Z_{2}^{I}$ duality symmetry $\l\to 1/\l$ \cite{gk}
accompanied by a reparamerization $\b\to \pi-\b$, $\cos\b\to -\cos\b$
and $\a\to -\a$ under which the background
fields and thus the CFT are invariant.
This is a parity-like symmetry (in the $\a$ direction) since if we do
not transform
$\a$ then $A_{a}\to -A_{a}$ and $A_{\g}\to A_{g}$.
There is another $Z_{2}^{II}$ duality which acts on $H$ as $H\to -H$.
This a charge conjugation symmetry since it changes the sign of the
gauge fields.
The combined transformation is a CP symmetry since it changes the
sign
of the Lorentz generator (in the third direction) $J^{3}+\bar J^{3}$.

Again here the deformed geometry is smooth for all values of $H,\l$
except at the boundaries $\l=0,\infty$ where the magnetic field turns
off and the three-sphere degenerates to the (classically) singular
geometry of the
$R\times SU(2)_{k}/U(1)$, \cite{gk}.

\section{Conformal Field Theory description of magnetic and
gravitational backgrounds}
\setcounter{equation}{0}

Our aim in this section is to define the deformation of the original
string ground state, that turns on magnetic fields and curvature, and
study the exact
spectrum. In particular we find the presence of instabilities of the
tachyonic type associated to such backgrounds.

We will focus on heterotic 4-D string ground states, described in
detail in the previous section, although the extension to type II
ground states is straightforward.
\def\q{{\cal Q}}

As mentioned in the introduction, the vertex operator which turns on
a chromo-magnetic field background $B^{a}_{i}$ is
\be
V^{a}_{i}=(J^i+{1\over 2}\epsilon^{i,j,k}\psi^j\psi^k)\bar J^{a}
\ee
This vertex operators is of the current-current type. In order for
such perturbations to be marginal (equivalently the background to
satisfy the string equations of motion) we need to pick a single
index $i$, which we choose to be $i=3$ and need to restrict the gauge
group index $a$ to be in the Cartan of the
gauge group.
We will normalize the antiholomorphic currents $\bar J^a$ in each
simple or U(1) component $G_{i}$ of the gauge group $G$ as
\be
\langle \bar J^a(\bar z)\bar J^b(0)\rangle ={k_{i}\over
2}{\delta^{ab}\over \bar z^2}\label{norm}
\ee
With this normalization, the field theory gauge coupling is
$g_{i}^2=2/k_{i}$.
Thus the most general (marginal) chromo-magnetic field is generated
from the following vertex operator
\be
V_{magn}={(J^3+\psi^1\psi^2)\over \sqrt{k+2}}{{\vec F_{i}}\cdot{\bar
{\vec J_{i}}}\over \sqrt{k_{i}}}\label{magn}
\ee
where the index $i$ labels the simple or $U(1)$ components $G_{i}$ of
the gauge group and $\bar{\vec J_{i}}$ is a $r_{i}$-dimensional
vector of currents in the Cartan of the group
$G_{i}$ ($r_{i}$ is the rank of $G_{i}$).
The repeated index $i$ implies summation over the simple components
of the gauge group.

We would like to obtain the exact one-loop partition function in the
presence of such perturbation. Since this is an abelian
current-current perturbation,
the deformed partition function can be obtained by an
$O(1,\sum_{i}r_{i})$ boost
of the charged lattice of the undeformed partition function, computed
in the previous section.

We will indicate the method in the case where we turn on a single
magnetic field $F$, a gauge group factor with central element
$k_{g}$, in which case
\be
V_{F}= F{(J^3+\psi^1\psi^2)\over \sqrt{k+2}}{\bar J\over
\sqrt{k_{g}}}
\ee
Let us denote by $\q$ the zero mode of the holomorphic helicity
current
$\psi^1\psi^2$,  $\pq$ the zero mode of the antiholomorphic current
$\bar J$ and $I,\bar I$ the zero modes of the $SU(2)$ currents
$J^{3},\bar J^{3}$ respectively.
Then, the relevant parts of $L_{0}$ and $\bar L_{0}$ are
\be
L_{0}={\q^2\over 2}+{I^2\over k}+\cdots\;\;\;,\;\;\;\bar
L_{0}={\pq^2\over k_{g}}+\cdots\label{k1}
\ee
We will rewrite $L_{0}$ as
\be
L_{0}={(\q+I)^2\over k+2}+{k\over 2(k+2)}\left(\q-{2\over
k}I\right)^2+\cdots
\ee
where we have separated the relevant supersymmetric zero mode $\q+I$
and its
orthogonal complement $\q-2I/k$ which will be a neutral spectator to
the perturbing process.
What remains to be done is an $O(1,1)$ boost that mixes the
holomorphic current
$\q+I$ and the antiholomorphic one $\pq$.
This is straightforward with the result
\be
L_{0}'= {k\over 2(k+2)}\left(\q-{2\over k}I\right)^2+\left(\cosh x
{\q+I\over \sqrt{k+2}}+\sinh x {\pq\over
\sqrt{k_{g}}}\right)^2+\cdots
\label{k2}
\ee
\be
\bar L_{0}'= \left(\sinh x {\q+I\over \sqrt{k+2}}+\cosh x {\pq\over
\sqrt{k_{g}}}\right)^2+\cdots\label{k3}
\ee
where $x$ is the parameter of the $O(1,1)$ boost.
Thus we obtain from (\ref{k2}), (\ref{k3}) the change of $L_{0}$,
$\bar L_{0}$ as
\be
\delta L_{0}\equiv L_{0}'-L_{0}=\delta \bar L_{0}\equiv   \bar
L_{0}'-\bar L_{0}=
F{(\q+I)\over \sqrt{k+2}}{\pq\over \sqrt{k_{g}}}+{\sqrt{1+F^2}-1\over
2}\left[
{(\q+I)^2\over k+2}+{\pq^2\over k_{g}}\right]
\ee
where we have identified
\be
F\equiv \sinh (2x)
\ee
we are now able to compute with the more general marginal
perturbation
which is a sum of the general magnetic perturbation (\ref{magn})
and the gravitational perturbation
\be
V_{grav}= \rt {(J^3+\psi^1\psi^2)\over \sqrt{k+2}}{\bar J^3\over
\sqrt{k}}
\ee
The only extra ingredient we need is an $O(1+\sum_{i}r_{i})$
transformation to mix
the antiholomorphic currents.
Thus, we obtain
\be
\delta L_{0}=\delta \bar L_{0}=\left[{\rt\bar I\over \sqrt{k}}+{{\vec
F}_{i}\cdot {\bar {\vec {\cal P}}}_{i}\over
\sqrt{k_{i}}}\right]{(\q+I)\over \sqrt{k+2}}+\label{gener}
\ee
$$+{\sqrt{1+\rt^2+{\vec F}_{i}\cdot {\vec F}_{i}}-1\over 2}\left[
{(\q+I)^2\over k+2}+(\rt^2+{\vec F}_{i}\cdot {\vec F}_{i})^{-1}\left(
{\rt\bar I\over \sqrt{k}}+{{\vec F}_{i}\cdot {\bar {\vec {\cal
P}}}_{i}\over \sqrt{k_{i}}}\right)^2\right]
$$

{}From now on we focus in the case where we have a single
chromo-magnetic field $F$ as well as the curvature perturbation
$\rt$.
Then (\ref{gener}) simplifies to

\be
\delta L_{0}=\delta \bar L_{0}=
\left[\rt{\bar I\over \sqrt{k}}+F{\pq\over
\sqrt{k_{g}}}\right]{(\q+I)\over \sqrt{k+2}}+\label{gener2}
\ee
$$+{\sqrt{1+\rt^2+F^2}-1\over 2}\left[
{(\q+I)^2\over k+2}+(\rt^2+F^2)^{-1}\left(
\rt{\bar I\over \sqrt{k}}+F{\pq\over \sqrt{k_{g}}}\right)^2\right]
$$

Eq. (\ref{gener2}) can be written in the following form which will be
useful
in order to compare with the field theory limit
\be
\delta L_{0}={1+\sqrt{1+F^2+{\cal R}^2}\over 2}\left[{(\q+I)\over
\sqrt{k+2}}+{1\over 1+\sqrt{1+F^2+{\cal R}^2}}\left(\rt{\bar I\over
\sqrt{k}}
+F{\pq\over \sqrt{k_{g}}}\right)\right]^2
\ee
$$-{(\q+I)^2\over k+2}
$$
and for $\rt=0$ as
\be
\delta L_{0}={1+\sqrt{1+F^2}\over 2}\left[{(\q+I)\over
\sqrt{k+2}}+{F\over 1+\sqrt{1+F^2}}{\pq\over \sqrt{k_{g}}}\right]^2
-{(\q+I)^2\over k+2}\label{k10}
\ee

Eq. (\ref{zero}) along with (\ref{gener}) provide the complete and
exact
spectrum of string theory in the presence of the chromo-magnetic
fields $\vec F_{i}$ and curvature $\rt$.
We will analyse first the case of a single magnetic field $F$ and use
(\ref{k10}). Let $L_{0}=M_{L}^2$ and $\bar L_{0}=M_{R}^2$.
Then

\be
M^2_{L}=-{1\over 2}+{\q^2\over 2}+{1\over
2}\sum_{i=1}^{3}\q_{i}^2+{(j+1/2)^2-(\q+I)^2\over
k+2}+E_{0}+\label{k11}
\ee
$$+{1+\sqrt{1+F^2}\over 2}\left[{(\q+I)\over \sqrt{k+2}}+{F\over
1+\sqrt{1+F^2}}{\pq\over \sqrt{k_{g}}}\right]^2
$$
\be
M^2_{R}=-1+{\pq^2\over k_{g}}+{(j+1/2)^2-(\q+I)^2\over
k+2}+\bar E_{0}+
\ee
$$+{1+\sqrt{1+F^2}\over 2}\left[{(\q+I)\over \sqrt{k+2}}+{F\over
1+\sqrt{1+F^2}}{\pq\over \sqrt{k_{g}}}\right]^2
$$
where, the $-1/2$ is the universal intercept in the N=1 side,
$\q_{i}$
are the internal helicity operators (associated to the internal
left-moving fermions),
$E_{0},\bar E_{0}$ contain the oscillator contributions as well as
the internal
lattice
(or twisted) contributions, and
$j=0,1,2,\cdots,k/2$\footnote{Remember that $k$ is an even integer
for $SO(3)$.}, $j\geq |I|\in Z$.
We can see here another reason for the need of the SO(3)
projection. We do not want half integral values of $I$ to change the
half-integrality of the spacetime helicity $\q$.
Since for physical states $L_{0}=\bar L_{0}$ it is
enough to look
at $M_{L}^2$ which in our conventions is the side that has
$N=1$ superconformal symmetry.

Let us consider first at how the low lying spectrum of space-time
fermions is modified.
For this we have to take $\q=\q_{i}=\pm 1/2$.
Then $M_{L}^2$ can be written as a sum of positive factors,
$E_{0}\geq 0$,
$(j+1/2)^2\geq (\pm 1/2+I)^2$ and
\be
 {1+\sqrt{1+F^2}\over 2}\left[{(\q+I)\over \sqrt{k+2}}+{F\over
1+\sqrt{1+F^2}}{\pq\over \sqrt{k_{g}}}\right]^2
\geq 0\label{k14}
\ee
Thus fermions cannot become tachyonic and this a good consistency
check for our spectrum since a ``tachyonic" fermion is a ghost.
This argument can be generalized to all spacetime fermions in the
theory.

Bosonic states can become tachyonic though, but for this to happen,
as in field theory they need to have non-zero helicity.
Since $(j+1/2)^2\geq I^2$ and $E_{0}\geq 0$, a state needs a non-zero
value for $\q$ and the minimum possible value for $\q_{i}^2$
(consistent with the GSO projection) as well as $E_{0}=0$ in order to
have a chance to
become tachyonic.
Also we need $j=\pm I$ and $\q I$ positive.
For such states, imposing $L_{0}=\bar L_{0}$ we obtain
\be
\q^2-{2\over k_{g}}\pq^2+1=2\bar E_{0}\geq 0\label{k16}
\ee
and thus the minimal value for $M_{L}^2$ can be written as
\be
M^2_{min}={\q^2-1\over 2}+{(|I|+1/2)^2-(\q+I)^2\over k+2}+
{1+\sqrt{1+F^2}\over 2}\left[{(\q+I)\over \sqrt{k+2}}+{F\over
1+\sqrt{1+F^2}}{\pq\over \sqrt{k_{g}}}\right]^2
\label{k13}
\ee
and due to Eq. (\ref{k14}) and the fact that $2I\leq k$ we obtain
\be
k|\q|(|\q|-2)\leq 3/2\;\;\;,\;\;\;|\q|=1,2,\cdots\label{k24}
\ee
\be
2|I|\geq {-k\q^2+k+3/2\over 1-2|\q|}\label{k25}
\ee
which imply that either $|\q|=1$ and $|I|=0,1,\cdots,k/2$, or
$|\q|=2$ and $|I|=k/2$, provided $k>0$.
However, due to the GSO projection, $\q$ mast be an odd integer.
Thus, for $k>0$ instabilities are due to helicity $\pm 1$ particles.

Let us introduce the variables
\be
H={F\over \sqrt{2}(1+\sqrt{1+F^2})}\;\;\;,\;\;\;e=\sqrt{2\over
k_{g}}\pq
\ee
$H$ is the natural magnetic field from the $\sigma$-model point of
view (see
section 3) and $e$ is the charge.
Notice that while $F$ varies along the whole real line, $|H|\leq
1/\sqrt{2}$. At $H_{max}=1/\sqrt{2}$ we can see from (\ref{k11}) that
there is an infinite number of states whose mass becomes zero, so it
is a decompactification boundary.

Eq. (\ref{k16}) can be rewritten as
\be
e^2\leq \q^2+1\label{k17}
\ee
Then, there are tachyons provided
\be
{1\over 1-2H^2}\left({(\q+I)\over
\sqrt{k+2}}+eH\right)^2+{\q^2-1\over 2}+{(|I|+1/2)^2-(\q+I)^2\over
k+2}\leq 0\label{tach}
\ee
For (\ref{tach}) to have solutions we must have
\be
e^2\geq \q^2-1+2{(|I|+1/2)^2\over k+2}\label{k18}
\ee
which along with (\ref{k17}) implies that $(|I|+1/2)^2\leq k+2$.
It is not difficult to see that the first instability sets in,
induced
from the $I=0$ states.
There is also an upper critical magnetic field beyond which no state
is tachyonic. This is obtained by considering the largest possible
value for $|I|$
(compatible with $(|I|+1/2)^2\leq k+2$).
We will leave the charge free for the moment although there are
certainly constraints on it depending on the gauge group.
For example for the $E_{6}$ or $E_{8}$ groups we have
$e^2_{min}=1/4$, and for all realistic non-abelian gauge groups
$e_{min}={\cal O}(1)$.
For toroidal $U(1)$'s however $e_{min}$ can become arbitrarily small
by tuning the parameters of the torus. Note however that in any case
for the potential tachyonic states with $|\q|=1$ the charge must
satisfy
\be
{1\over 2(k+2)} \leq e^2 \leq 2\label{k20}
\ee

Thus for $|\q|=1$ we obtain the presence of tachyons provided that
\be
H^{\rm crit}_{\rm min}\leq |H| \leq H^{\rm crit}_{\rm max}
\ee
with
\be
H^{\rm crit}_{\rm min}={\mu\over |e|} {1-{\sqrt{3}\over
2}\sqrt{1-{1\over 2}\left({\mu\over e}\right)^2}\over
1+{3\over 2}\left({\mu\over e}\right)^2}
\ee
\be
H^{\rm crit}_{\rm max}={\mu\over |e|}{J+1+ \sqrt{\left(J+{3\over
4}\right)\left(1-2\left(J+{1\over 2}\right)^2{\mu^2\over
e^2}\right)}\over
1+\left(2J+{3\over 2}\right){\mu^2\over e^2}}
\label{q1}
\ee
where
\be
J={\rm integral ~~part~~of~~~}-{1\over 2}+{|e|\over \sqrt{2}\mu}
\ee
We have also introduced the IR cutoff scale $k+2=1/\mu^2$.

We note that for small $\mu$ and $|e|\sim {\cal O}(1)$ $H^{\rm
crit}_{\rm min}$ is of order ${\cal O}(\mu)$.
However $H^{\rm crit}_{\rm max}$ is below $H_{\rm max}=1/\sqrt{2}$ by
an amount of order ${\cal O}(\mu)$.
Thus for small values of $H$ there are no tachyons until a critical
value $H^{\rm crit}_{\rm min}$ where the theory becomes unstable. For
$|H|\geq H^{\rm crit}_{\rm max}$ the theory is stable again till the
boundary $H=1/\sqrt{2}$.
It is interesting to note that if there is a charge in the theory
with the value $|e|=\sqrt{2}\mu$ then $H^{\rm crit}_{\rm
max}=1/\sqrt{2}$ so there
is no region
of stability for large magnetic fields.
For small $\mu$ there are always charges satisfying (\ref{k20}) which
implies that there is always a magnetic instability.
However even for $\mu={\cal O}(1)$ the magnetic instability is
present for standard gauge groups that have been considered in string
model building (provided it has charged states in the perturbative
spectrum).

The behavior above should be compared to the field theory behavior
(\ref{class}). There we have an instability provided there is a
particle with $gS\geq 1$. Then the theory is unstable for
\be
|H|\geq {M^2\over |e|(gS-1)}
\ee
where $M$ is the mass of the particle (or the mass gap).
However there is no restauration of stability for large values of
$H$.
This happens in string theory due to the backreaction of gravity.
There is also another difference. In field theory $H_{crit}\sim
\mu^2$
while in string theory $H_{crit}\sim \mu M_{\rm planck}$ where we
denoted by
$\mu$ the mass gap in both cases and $M^2_{str}=1/\alpha'g^2_{\rm
string}$.

We will also study the special case $k=0$, which was left out from
the analysis above.
This corresponds to a strongly curved spacetime (the curvature of
$S^3$ is of the order of the string scale). We know of course from
the CFT that for $k=0$
the $S^3$ decouples (only the ground state is left).
This is a non-critical string theory since from the bosonic part
of 4-d, only the Liouville field survives.
Moreover, $H$ loses its meaning as a magnetic field (since it couples
only to the helicity operator).
In this case all (odd) helicity states can become tachyonic and we
obtain
an instability for
\be
H_{min}^{k=0}\leq H^{k=0} \leq H_{max}^{k=0}
\ee
with
\be
H^{k=0}_{\rm max}=H_{\rm max}={1\over \sqrt{2}}
\ee
\be
H^{k=0}_{\rm min}={1\over \sqrt{2}}{|e|-\sqrt{3(e^2-1/4)/4}\over
e^2+3/4}
\ee
The first tachyonic instability related to $H^{k=0}_{\rm min}$ is
induced by
$|\q|=1$ states. The theory never becomes stable again since for all
$H\leq H_{\rm max}$ there are tachyonic states for arbitrary high
values of $|\q|$.

The analysis above applies to magnetic fields embedded in non-abelian
gauge groups, not broken by the conventional Higgs effect.
We will also consider however broken non-abelian gauge groups.
Consider the internal CFT containing a circle of radius R.
For arbitrary values of $R\not= 1$ there is a U(1) gauge symmetry
which is enhanced at $R=1$ to $SU(2)$.
The $W^{\pm}$ bosons have masses proportional to $(R-1/R)^2$ and
become massless at $R=1$.

In such a case we will again consider states with $\Q=1$, $\Q_{i}=0$,
$E_{0}=(R-1/R)^2/4$ and $\pq/\sqrt{k_{g}}=(R+1/R)/2$ in (\ref{k11}).
The condition for the W-bosons becoming tachyonic is
\be
(1-2H^2)\left[{1\over 4}\left(R-{1\over R}\right)^2-\left(|I|+{3\over
4}\right)
\mu^2\right]+\left[(|I|+1)\mu+{H\over \sqrt{2}}\left(R+{1\over
R}\right)\right]^2\leq 0
\label{u1}\ee
It is obvious that the first factor has to be negative so that there
tachyons provided
\be
{1\over 4\mu^2}\left(R-{1\over R}\right)^2-{3\over 4}\leq |I|\leq
{1\over 2\mu^2}-1
\ee
from which we obtain
\be
{4-\mu^2-\sqrt{(4-\mu^2)^2-4}\over 2}\leq R^2 \leq
{4-\mu^2+\sqrt{(4-\mu^2)^2-4}\over 2}
\ee
Note that this condition is duality invariant.
Again here we have two critical values for the magnetic field as
before that can be computed from (\ref{u1}).
However there is no instability in the flat limit $\mu\to 0$ unlike
the field theory case (\ref{sta}) due to the gravitational back
reaction.

Let us now study the gravitational perturbation. Using (\ref{gener2})
the mass formula is (in analogy with (\ref{k11})
\be
M^2_{L}=-{1\over 2}+{\q^2\over 2}+{1\over
2}\sum_{i=1}^{3}\q_{i}^2+{(j+1/2)^2-(\q+I)^2\over
k+2}+E_{0}+\label{k22}
\ee
$$+{1+\sqrt{1+\rt^2}\over 2}\left[{(\q+I)\over \sqrt{k+2}}+{\rt\over
1+\sqrt{1+\rt^2}}{\bar I\over \sqrt{k}}\right]^2
$$
Introducing the $\sigma$-model variable
\be
\l=\sqrt{\rt +\sqrt{1+\rt^2}}\;\;\;,\;\;\;{1\over \l}=\sqrt{-\rt
+\sqrt{1+\rt^2}}
\ee
(\ref{k22}) becomes
\be
M^2_{L}=-{1\over 2}+{\q^2\over 2}+{1\over
2}\sum_{i=1}^{3}\q_{i}^2+{(j+1/2)^2-(\q+I)^2\over
k+2}+E_{0}+\label{k23}
\ee
$$+{1\over 4}\left[\left(\l+{1\over \l}\right){(\q+I)\over
\sqrt{k+2}}+\left(\l-{1\over \l}\right){\bar I\over
\sqrt{k}}\right]^2
$$
Eqs (\ref{k24}) and (\ref{k25}) are still applicable here, which
means
that we have to examine only $|\q|=1$ and $|I|=0,1,\cdots,k/2$, or
$|\q|=2$ and $|I|=k/2$.
Again $j=|I|$ and $I\q >0$.
Due to the $\l\to 1/\l$ duality we will restrict ourselves to the
region $\l\leq 1$.

Thus, the condition for existence of tachyons is
\be
{1\over 4}\left[\left(\l+{1\over \l}\right){(\q+I)\over
\sqrt{k+2}}+\left(\l-{1\over \l}\right){\bar I\over
\sqrt{k}}\right]^2+
{\q^2-1\over 2}+{(|I|+1/2)^2-(\q+I)^2\over k+2}\leq 0\label{tach2}
\ee
In order that (\ref{tach2}) have solutions we must have
\be
\left[{\q^2-1\over 2}+{(|I|+1/2)^2-(\q+I)^2\over k+2}\right]
\left[{\q^2-1\over 2}+{(|I|+1/2)^2-\bar I^2\over k}\right]\geq 0
\label{k26}
\ee
The first factor was arranged already to be negative so we must
ensure that the second factor is also negative. This is impossible
for $|\q|=2$. Thus we are
left with $|\q|=1$ and
\be
|\bar I|\geq \sqrt{k\over k+2}\left(|I|+{1\over 2}\right)
\label{k27}
\ee

Thus the state with quantum numbers $(I,\bar I)$ satisfying
(\ref{k27}) becomes tachyonic when
\be
\l^2_{\rm min}\leq \l^2 \leq \l^2_{\rm max}
\ee
with
\be
\l^2_{\rm max}={{\bar I^2\over k}-{I^2-1/2\over
k+2}+\sqrt{{(I+3/4)\over
k+2}\left(
{\bar I^2\over k}-{(I+1/2)^2\over k+2}\right)}\over \left({I\over
\sqrt{k+2}}
+{\bar I\over \sqrt{k}}\right)^2}
\ee
\be
\l^2_{\rm min}={{\bar I^2\over k}-{I^2-1/2\over
k+2}-\sqrt{{(I+3/4)\over
k+2}\left(
{\bar I^2\over k}-{(I+1/2)^2\over k+2}\right)}\over \left({I\over
\sqrt{k+2}}
+{\bar I\over \sqrt{k}}\right)^2}
\ee
For large $k$, $\l_{\rm max}$ approaches one, however at the same
time the
instability region shrinks to zero so that in the limit
$\l=1,k=\infty$ flat space is stable.

\section{The Flat Space Limit}\setcounter{equation}{0}

As mentioned earlier, in the limit $k\to \infty$ the 4-d space
becomes flat
(with zero dilaton). We would like to understand the nature of the
magnetic
fields in this limit.

As a warm-up we will describe first (in the context of field theory)
the case of a constant magnetic field in flat space as a limit of a
monopole field of a
two-sphere in the limit that the radius of the sphere becomes large.
Let $g$ be the strength of the monopole field. Then
\be
\vec B_{monopole}={g}{{\vec r}\over r^3}\label{33}
\ee
We have the Dirac quantization condition for $g$ in terms of the
elementary
electric charge $e$: $eg=N$ where $N$ is an arbitrary positive
integer
or half-integer.
Let us now consider a charged spinless particle of charge $e$
constrained to move on a two-sphere around the origin, of radius $R$.
The (non-relativistic\footnote{The relativistic case is similar up to
the zero point shift $m_{0}$ of the energy, a scaling of the Landau
spectrum by $1/m_{0}$ and ${\cal O}(m_{0}^{-3})$ relativistic
corrections.}) spectrum of such a particle is known \cite{col}:
\be
\Delta E_{j}={1\over R^2}\left[j(j+1)-N^2\right]
\label{34}\ee
where $j=N,N+1,\cdots$ and $N=eg\in Z/2$.
For each $j$ there are $2j+1$ degenerate states.
If we define $n=j-N$ then
\be
\Delta E_{n}={1\over
R^2}\left[n(n+1)+N(2n+1)\right]\;\;\;,\;\;\;n=0,1,\cdots\label{35}
\ee
We would like now to take $R\to \infty$. There are two possible
limits to consider.
The first is the limit where the magnetic flux per unit area is
finite.
Since the area of the sphere becomes infinite in the flat limit we
will have
to take the monopole strength to $\infty$ as $g=HR^2+{\cal O}(R)$
where $H$ is the flat space magnetic field.
Then $eg=N=eHR^2+{\cal O}(R)$ and
\be
\Delta E^{2d-flat\;\;space}_{n}=eH(2n+1)+{\cal
O}(R^{-1})\;\;\;,\;\;\;n=0,1,\cdots
\label{36}
\ee
We have thus recovered the usual formula where $n$ labels the Landau
levels
and we should remember that each Landau level is infinitely
degenerate corresponding to different values of the angular momentum
in the direction perpendicular to the plane.

The other limit is to keep the monopole strength fixed. In this case
we end up with a zero flat space magnetic field and continuous
spectrum $E=p^2$ corresponding, not surprisingly, to a free particle
in 2-d.

Let us consider now (again in the context of field theory) a magnetic
field
on a three-sphere of radius $R$. In Euler angles:
\be
A_{\a}=H\cos\b\;\;,\;\;A_{\b}=0\;\;,\;\;A_{\g}=H
\ee
which is exactly the same as the stringy background (\ref{10}) we
have found earlier.
We will find again the energy spectrum of a particle of electric
charge $e$
moving on $S^3$. The Hamiltonian  is as usual
\be
\hat {\bf H}={1\over \sqrt{{\rm det}
G}}\left(\partial_{\mu}-ieA_{\mu}\right)
\sqrt{{\rm det}G}G^{\mu\nu}\left(\partial_{\nu}-ieA_{\nu}\right)
\label{38}
\ee
where $G_{\mu\nu}$ is the round metric on $S^{3}$. Notice that it is
different
from the stringy metric (\ref{8}),(\ref{9}) which contains the
gravitational
backreaction.
It is straightforward to work out the spectrum of $\hat {\bf H}$ with
the result:
\be
\Delta E_{j,m}={1\over R^2}\left[j(j+1)-m^2+(eH-m)^2\right]\label{39}
\ee
where for $SO(3)$ $j\in Z$ and $-j\leq m\leq j$.
We can always parametrize $j,m$ as $j=|m|+n$ with $|m|=0,1,\cdots$
and $n=0,1,2,\cdots$ so
\be
\Delta E_{n,m}={1\over
R^2}\left[n(n+1)+|m|(2n+1)\right]+\left({eH-m\over R}\right)^2
\label{40}\ee
In order to take the flat space limit and recover Landau levels we
have to
scale $eH=e\tilde H R^2+\kappa R+{\cal O}(1)$ and $m=e\tilde H
R^2+(p_{3}+\kappa) R+{\cal O}(1)$.
Then we obtain from (\ref{40})
\be
\Delta E_{n,p_{3}}=e\tilde H(2n+1)+p_{3}^2+{\cal O}(R^{-1})
\label{41}\ee
which is the standard Landau spectrum in 3-d flat space.
This reproduces (\ref{class}) for spinless particles, $S=0$.

In the discussion above  we did not include the gravitational
backreaction since we had a round metric for $S^{3}$.
This is what we are going to do now. We will start from the
background
(\ref{8}),(\ref{9}),(\ref{10}) and compute the energy eigenvalues of
the
(field theory) Hamiltonian given by (\ref{38}).
This is straightforward with the answer:
\be
\Delta E_{j,m}={1\over R^2}\left[j(j+1)-m^2+{(eHR-m)^2\over
(1-2H^2)}\right]\label{42}
\ee
and after parametrizing again $j=|m|+n$ with $|m|=0,1/2,1,\cdots$
and $n=0,1,2,\cdots$ we obtain
\be
\Delta E_{n,m}={1\over
R^2}\left[n(n+1)+|m|(2n+1)\right]+\left({eHR-m\over
R\sqrt{1-2H^2}}\right)^2
\label{43}\ee
Notice that the only difference from (\ref{40}) (apart from the
different scaling of H which is a convention) is the extra $1-2H^2$
factor in the denominator of the last term.
This factor however makes the flat limit quite different.
In fact we can see that in order to have Landau levels we have to
take
$m\sim {\cal O}(R^2)$, in which case we are forced to have from the
last term
that $H\sim {\cal O}(R)$ in which case the denominator gives a
negative contribution.
This is obvious from the fact that since there is a maximal value for
H we cannot take it to scale as the radius R.
So Landau levels disappear from the low energy spectrum, and with
$m=pR+{\cal O}(1)$ we obtain in the limit $R\to\infty$
\be
\Delta E_{p}={(p-eH)^2\over 1-2H^2}+{\cal O}(R^{-1})
\label{44}
\ee
This implies that the flat space limit is quite different from
standard field theory.
We have already seen in the previous section that $W$ bosons do not
become tachyonic in the flat limit.

The field theory spectrum parallels the exact string spectrum in the
presence
of a magnetic field.
The correct identification there is
\be
R\to k+2\;\;\;,\;\;\;m\to Q+J^3\;\;\;,\;\;\; e\to \sqrt{2\over
k_{g}}\bar Q
\label{450}
\ee
\be
H\to {F\over \sqrt{2}(1+\sqrt{1+F^2})}={1\over
2\sqrt{2}}\left[F-{F^3\over 4}+
{\cal O}(F^5)\right]
\ee
In terms of the CFT variable $F$ the maximum magnetic field
$H_{max}=1/\sqrt{2}$ corresponds to the limit $F\to \pm\infty$.
As we have seen already the tachyonic instabilities appear before the
magnetic field reaches its maximum value.

We can now discuss the spectra of particles with spin.
For particles that inherit their spin from the helicity operator
(this includes massless fermions, and heterotic gauge fields) we can
set $S=Q$
and using the string identification (\ref{450}) we obtain the
following spectrum
\be
\Delta E_{j,m,S}={1\over k+2}\left[j(j+1)-(m+S)^2+{(eHR-m-S)^2\over
(1-2H^2)}\right]\label{422}
\ee
which to linear order in the magnetic field becomes
\be
\Delta E_{j,m,S}={j(j+1)\over k+2}-{2eH\over \sqrt{k+2}}(m+S)+{\cal
O}(H^2)
\ee
which indicates the possibility of tachyons.
The existence of tachyonic modes with non-zero spin was verified in
section 4.

In a similar manner we can compute the (scalar) field theory spectrum
in the combined magnetic and gravitational background
(\ref{11})-(\ref{20})
\be
\Delta E_{j,m,\bar m}= {1\over
R^2}\left[j(j+1)-m^2+{(2eHR-(\l+1/\l)m-(\l-1/\l)\bar m)^2\over
4(1-2H^2)}\right]
\label{46}
\ee
where $j\in Z$ and $-j\leq m,\bar m\leq j$.
(\ref{46}) reduces to (\ref{42}) when $\l=1$.
Here we see that we can adjust the extra modulus $\l$ in order to
obtain Landau levels in the large volume limit.
However the coefficient is not related to the magnetic field H, since
the cancelations in the last term are due to a tuning of the modulus
$\l$ which takes large (or small via duality) values. The
interpretation of this limit is the following. Let us first take
$H=0$ since it is not relevant in this limit.
{}From the point of view of the coset space $SU(2)/U(1)$,
the SU(2) WZW model can be viewed as a Dirac monopole on $S^2$,
\cite{bk}.
Thus at $\l=1$ we can write (\ref{46}) in the form
\be
\Delta E_{j,m,\bar m}= {1\over R^2}\left[j(j+1)-m^2\right]+{m^2\over
R^2}={j(j+1)\over R^2}
\label{466}\ee
In (\ref{466}) the piece $j(j+1)-m^2$ of the energy is the standard
spectrum of charged particles in the presence of the monopole and the
additional $m^2$ is coming from the Kaluza-Klein masses of the
charged
modes.
The states with non-trivial $\bar m$ are not charged with respect to
the monopole and thus do not contribute to the energy.
When we perturb $\l$ away from 1 we can suppress the Kaluza-Klein
masses
and thus we can have a limit similar to that of Eq. (\ref{36}).

If we include all higher order corrections in $\alpha'$ and identify
$R^2\to k+2$ then (\ref{46}) becomes

\be
\Delta E_{j,m,\bar m}= {1\over
k+2}\left[j(j+1)-m^2\right]+{\left(2\sqrt{k+2}eH-\left(\l+{1\over
\l}\right)m-\left(\l-{1\over \l}\right)\sqrt{(1+2/k)}\bar
m\right)^2\over 4(k+2)(1-2H^2)}
\label{47}
\ee
Eq. (\ref{47}) matches the string theory spectrum with the following
identifications
\be
m\to Q+J^3\;\;\;,\;\;\;  e\to \sqrt{2\over k_{g}}\pq
\;\;\;,\;\;\;\bar m\to \bar J^3
\ee
\be
H^2={1\over 2}\;\;{F^2\over F^2+2\left(1+\sqrt{1+F^2+{\cal R}^2}\right)}
={F^2\over
8}\left[1+{\cal O}(F^2,{\cal R}^2)\right]
\ee
\be
\l^2= {1+\sqrt{1+F^2+{\cal R}^2}+{\cal R}\over
1+\sqrt{1+F^2+{\cal R}^2}-{\cal R}}
=1+{\cal R}+
{\cal O}(F^2,{\cal R}^2)
\ee
\vskip 1cm

\section{Conclusions and Further Comments}

We have studied a class of magnetic and gravitational backgrounds in
closed superstrings and their associated instabilities.
Our starting point are superstring ground states with an adjustable
mass
gap $\mu^2$ \cite{kk}. We have described in detail how to construct
them starting from any four-dimensional ground state of the string,
by giving appropriate expectation values to the graviton
antisymmetric tensor and dilaton.
In such ground states all gauge symmetries are spontaneously broken.

Exact magnetic and gravitational solutions can then be constructed in
such ground states as exactly marginal perturbations of the
appropriate CFTs.
In the magnetic case, there is  a monopole-like magnetic field on
$S^3$.
The gravitational backreaction squashes mildly the $S^3$ keeping
however an $SO(3)$ symmetry.
We have calculated the exact spectrum as a function of the magnetic
field.
The first interesting observation is that, unlike field theory, there
is a
maximum value for the magnetic field $\sim M_{\rm planck}^2$.
At this value the part of the spectrum that couples to the magnetic
field
becomes infinitely massive.

We find magnetic instabilities in such a background.
In particular, for $H\sim {\cal O}(\mu M_{\rm planck})$ there is a
magnetic instability,
driven by helicity-one states that become tachyonic.
The critical magnetic field scales differently from the field theory
result, due the different mechanism of gauge symmetry breaking.

We also find that, unlike field theory, the theory becomes stable
again for
strong magnetic fields of the order $\sim {\cal O}(M^{2}_{\rm
planck})$.

Similar behavior is found for the gravitational perturbation.
Here again there is an intermediate region of instability in the
perturbing parameter.

Such instabilities could be relevant in cosmological situations, or
in black hole evaporation.
In the cosmological context, there maybe solutions where one has time
varying
long range magnetic fields. If the time variation is adiabatic, then
there might be
a condensation which would screen and localize the magnetic fields.
Results on such cosmological solutions will be reported elsewhere.
Also, instabilities can be used as (on-shell) guides to find the
correct vacuum of string theory.
Our knowledge in that respect is limited since we do not have an
exact description of all possible deformations
of a ground state in string theory.

Another subject of interest, where instabilities could be relevant is
Hawking radiation.
It is known in field theory that Hawking radiation has many common
features with production
of Schwinger pairs in the presence of a long range electric field.
In open string theory it was found, \cite{bp} that this rate becomes
infinite for a
$finite$ electric field , $E_{\rm crit}\sim M^2_{\rm string}$ (unlike
the field theory case) and this behavior is due
to $\alpha'$ corrections. Notice also that in the open string it is
$M_{\rm string}$ and not $M_{\rm planck}$ that is relevant due to the
absence of gravity.
It would be interesting to see if this behavior persists in the
presence of gravity
(which is absent to leading order in open strings)
by studying the effect in closed strings.
In fact we expect that gravitational effects will be important for
$E\sim M^2_{\rm planck}$. For small $g_{\rm string}$ however,
we can have
$M_{\rm string} << M_{\rm planck}$ so we expect a similar behavior as
in the case of open strings.
It is plausible that similar higher order corrections modify the
Hawking rate
in such a way that (some) black holes are unstable in string
theory.
Such a calculation seems difficult to perform with today's technology
but seems crucial to
the understanding of stringy black holes.

\centerline{\bf Acknowledgements}
We would like to thank L. Alvarez-Gaum\'e, S. Coleman, M. Porrati
and especially
J. Russo for discussions.
C. Kounnas was  supported in part by EEC contracts
SC1$^*$-0394C and SC1$^*$-CT92-0789.

\end{document}